%% file: IV_poststratification-applied.tex
\documentclass[12pt]{article}
\usepackage{amsmath}
\usepackage{amssymb}
\usepackage{amsthm}
\usepackage{graphicx,psfrag,epsf}
\usepackage{psfrag,epsf}
\usepackage{enumitem}
\usepackage{natbib}
\usepackage{url} 
\usepackage[titletoc,title]{appendix}
\usepackage{array}
\usepackage{bm}
\usepackage{color}
\usepackage{graphicx}
\usepackage{epstopdf}
\usepackage[letterpaper, portrait, margin=1in]{geometry}
\usepackage{lipsum}
\usepackage{listings}
\usepackage{fancyref}
\usepackage{float}
\usepackage[font=footnotesize,labelfont=bf]{caption}
\usepackage{color}	
\usepackage{lscape}
\usepackage{mathtools}

\usepackage{xr}
\usepackage{xcite}
\usepackage{caption}
\usepackage{subcaption}
\usepackage{booktabs}
\usepackage[normalem]{ulem}

\newtheorem{lemma}{Lemma}[section]
\newtheorem{theorem}{Theorem}[section]

\newtheorem{assumption}{Assumption}[section]
\newtheorem{result}{Result}[section]

\usepackage{setspace}

\newcommand*\ITT{\text{ITT}}
\newcommand*\CACE{\tau}

\title{Improving instrumental variable estimators with post-stratification}
\author{  Nicole E. Pashley\\Department of Statistics, Rutgers University \and Luke Keele \\Departments of Surgery and Biostatistics, University of Pennsylvania\and Luke W. Miratrix\\Graduate School of Education, Harvard University }

\begin{document}
\maketitle

\begin{abstract}
Experiments studying get-out-the-vote (GOTV) efforts estimate the causal effect of various mobilization efforts on voter turnout. However, there is often substantial noncompliance in these studies. A usual approach is to use an instrumental variable (IV) analysis to estimate impacts for compliers, here being those actually contacted by the investigators. Unfortunately, popular IV estimators can be unstable in studies with a small fraction of compliers. We explore post-stratifying the data (e.g., taking a weighted average of IV estimates within each stratum) using variables that predict complier status (and, potentially, the outcome) to mitigate this. We present the benefits of post-stratification in terms of bias, variance, and improved standard error estimates, and provide a finite-sample asymptotic variance formula. We also compare the performance of different IV approaches and discuss the advantages of our design-based post-stratification approach over incorporating compliance-predictive covariates into the two-stage least squares estimator. In the end, we show that covariates predictive of compliance can increase precision, but only if one is willing to make a bias-variance trade-off by down-weighting or dropping strata with few compliers. By contrast, standard approaches such as two-stage least squares fail to use such information. We finally examine the benefits of our approach in two GOTV applications.
\end{abstract}

\noindent%
\vfill

\doublespacing

\section{Introduction}

In United States elections, political parties often focus on voter mobilization, encouraging their partisans to vote via appeals made prior to an election. Voter mobilization contact is usually a simple reminder to vote, but may also convey information such as the day of the election and the location of polling places. A key question for political parties is what types of get-out-the-vote (GOTV) efforts (e.g., door-to-door canvasing, phone calls, texts, and so forth) are most effective. However, judging the effectiveness of GOTV methods is complicated by the fact that those voters most receptive to contact by political parties are already more likely to vote on election day. In ground-breaking research, \citet{Gerber:2000} used a randomized control trial (RCT) design to evaluate GOTV approaches, randomly allocating voters to different GOTV approaches and comparing voting behavior using public records. Gerber and Green spawned a new area of research based on using RCTs to identify best practices for increasing voter turnout rates \citep{Nickerson:2006,arceneaux2012get,green2017much,coppock2022does,mann2020negatively,Green:2003a,Gerber:2008}.
See \citet{green2019get,green2016voter,green2013field,Nickerson:2009} for overviews and meta-analyses of this literature.

Despite the merits of these experiments in terms of their ability to quantify causal quantities, GOTV RCTs are often subject to significant noncompliance. For example, \citet{Green:2003a} was designed to gauge the effectiveness of door-to-door canvassing by having volunteers knock on doors urging people to vote in an upcoming election in six cities (Bridgeport, Columbus, Detroit, Minneapolis, Raleigh, and St. Paul). In each city, voters in households were randomized to either receive face-to-face contact from local staffers, i.e. treatment, or were not contacted, i.e. control. In the study, using official lists of registered voters, voters were grouped into small geographic areas called turfs. Within each turf, voters were randomly assigned to treatment or control. Canvassers were then given the names and addresses of voters within each turf and instructed to only contact voters selected for treatment. However, only around 30\% of the voters assigned to treatment were actually contacted. In RCTs of this type, rates of contact often range between 10 to 30\% \citep{Green:2003a}.

When there is noncompliance, one strategy is to focus on the causal effect of the treatment assignment on the outcome, which is referred to as an intention-to-treat (ITT) analysis. However, there is also substantial interest in the causal effect of the treatment actually received. That is, we might wish to estimate the causal effect of actually being contacted rather than just the effect of being assigned to contact. When noncompliance is present, treatment assignment can be used as an instrumental variable (IV), which is a variable that affects exposure to treatment but does not directly affect the outcome \citep{AngImbRub96,Hernan:2006}. For a variable to be an instrument the following three core assumptions must hold: (1) the IV must have a nonzero effect on treatment exposure, (2) the IV must be randomly or as-if randomly assigned, (3) the IV must itself not have a direct effect on the outcome \citep{AngImbRub96}. If these assumptions hold in addition to a monotonicity assumption on the effect of assigning treatment on treatment receipt, an IV design provides a consistent estimate of the causal effect of the exposure on the outcome for so-called \emph{compliers} \citep[those who only take treatment upon encouragement,][]{AngImbRub96}, even in the presence of unobserved confounding between the treatment and the outcome.

In the original analysis of \cite{Green:2003a}, the authors first focused on ITT effects. They then used IV methods to estimate the effect of exposure on voter turnout. In RCT designs of this type, the IV assumptions are plausible. That is, (1) one can easily verify that treatment assignment causes treatment exposure, (2) the IV is randomly assigned by the study design, and (3) it is unlikely that being assigned to household contact has any direct effect on voting except through the exposure of a face-to-face appeal to vote. Unfortunately, even if the IV assumptions are met, if the instrument has only a small impact on the proportion of subjects who take treatment, such as with these GOTV experiments, the instrument is said to be weak \citep{Bound:1995,Staiger:1997}, and inference can be hard. In particular, with a weak instrument, IV estimates may be biased and the associated confidence intervals can have poor coverage. As such, analysts generally test for the presence of a weak instrument as an initial step in an IV analysis \citep{Stock:2005}. One interesting question is how analysts might improve GOTV and other studies where the instrument manages to pass a weak instrument test, but is still not strong. In this study, we focus on how to use baseline covariates (i.e., covariates measured pre-intervention or otherwise known to not be impacted by treatment) to improve IV estimates in applications and contexts such as these.

Specifically, we outline how post-stratification based on baseline covariates (hereafter referred to simply as ``covariates'') that are predictive of \emph{compliance type} can improve IV analyses. Generally, the literature has not focused on how to exploit covariates only predictive of compliance type, i.e., of being a complier or not. We show that covariates predictive of compliance type do not benefit an analysis the same way as covariates that improve precision. In particular, we demonstrate that classic IV estimation methods such as two-stage least squares (2SLS) provide minimal gains with complier-predictive covariates. We therefore propose a method based on post-stratification, where we stratify units based on baseline covariates, estimate a separate IV estimator within each stratum, and then take their average, weighting by the estimated number of compliers. If some strata are estimated to have zero compliers, we drop them from our overall estimate. The goal is to concentrate our compliers into a few strata where we can estimate impacts more easily. The strata with only a few compliers can be down-weighted, as they would be less relevant for the overall impact estimate.

Our proposal adds to the extensive literature on how to use covariates in IV contexts. For example, covariates can be used to bound effects when the exclusion restriction, a key IV assumption, does not hold \citep{mealli2013using,miratrix2018bounding}. Covariates also play a vital role in the principal scores literature \citep{ding2017principal, feller2017principal}, which relaxes the exclusion restriction with a weaker principal ignorability assumption. Covariates are also routinely used in Bayesian principal stratification to reduce model dependence and improve precision \citep{imbens1997bayesian, hirano2000assessing, mealli2012refreshing}.
\citet{schochet2024design} derives design-based finite-population central limit theorems for instrumental variable estimators adjusted for covariates through regression, focusing on strong instrument settings. \citet{ten2004causal} used covariates to predict compliance classes to target the intention-to-treat (ITT) estimand. Our proposed post-stratification methods also connect to, for a continuous IV case, matching methods where units different in terms of their exposure to the IV are matched on covariates \citep{Small:2008,baiocchi2012near,keele2016strong}. This matching IV literature proposes using an estimator similar to one of our post-stratified IV estimators. However, the goal in the matching IV literature is to mitigate bias from confounders in the observational setting and to match units that are far apart in terms of the instrument. We, on the other hand, focus on the use of post-stratification (which is more general than standard pairwise matching) to reduce variation, and we also discuss how modifications of the natural post-stratified IV estimator can lead to better precision gains.

Ideally, the post-stratification strategy would give a more precise estimator if the covariates used for stratification are predictive of the outcome or compliance, due to inherently more stable estimates within each group. Surprisingly, we find that although this estimator does take advantage of stratification variables predictive of outcome, it can fail to take advantage of covariates predictive of compliance type. Furthermore, most gains are driven by dropping empty strata, because of the precision boost from eliminating unstable IV estimators that are just contributing noise to the overall estimator. However, as we show, if we are not able to separate compliers and non-compliers cleanly enough to reliably drop any strata, our post-stratified estimator is identical to the ratio of a post-stratified estimate of the ITT effect and a post-stratified estimate of the compliance proportion, which in turn is very similar to the ineffective 2SLS estimator. As such, we explore other stratification estimators that more aggressively drop or down-weight those strata with few compliers to obtain further gains. Dropping or down-weighting strata does come at a cost of some additional bias, however, leaving us with a bias-variance tradeoff when using complier-predictive covariates.

In sum, we outline how post-stratification estimators can provide three potential benefits for impact estimation: (1) a reduction in the variance, (2) a reduction in the bias, and (3) a reduction in the variability of the estimated standard error. We first provide theoretical derivations for the behavior of this estimator. We then study the properties of the post-stratification IV estimator using a detailed simulation study. We find that post-stratification shows improvement in overall precision with covariates predictive of outcome, and that the versions that drop low weight strata can also achieve precision gains at the cost of a small amount of bias. We then apply these methods to two GOTV examples. We conclude with a discussion of when post-stratification IV estimators could be usefully employed and the implications of our findings for future GOTV study planning.

\section{The IV Framework}

We have $N$ units, and use the potential outcomes framework as introduced for RCTs in \citet{SplawaNeyman:1990ux}; see \citet{gerber2012field} for a more modern overview.
Under this framework, each unit has treatment indicator $Z_i \in \{0,1\}$ and associated potential outcomes under treatment and control, $Y_i(1)$ and $Y_i(0)$.
We also have indicators for actual treatment receipt when assigned to treatment $z$, $D_i(z) \in \{0,1\}$.
$D_i$ and $Y_i$ will be the observed values for unit $i$ after treatment assignment.
We also denote observed baseline covariates as $\bm X_i$.

Following the traditional Neyman-style causal inference framework, we assume the stable unit treatment value assumption (SUTVA) \citep{rubin_1980}.
Under SUTVA, both the potential treatments received $D(z)$ and potential outcomes $Y(z)$, $Y(z,d)$, for $z, d=0,1$, depend solely on the value $z$ of the instrument, and $d$ of the treatment for $Y(z,d)$, for each individual.
That is, there is only one version of the instrument and treatment, and $D_i(z)$, $Y_i(z)$, and $Y_i(z,d)$ are not affected by the value of $(Z_{i'}, D_{i'})$ for $i'\neq i$. These two components of SUTVA are often referred to as the consistency and no-interference assumptions, respectively. Next, we make the three core IV assumptions are \citep{AngImbRub96}:

\begin{assumption}\label{assump:iv} The three core IV assumptions:

\begin{enumerate}[wide=\parindent, label={\textbf{Part~\Alph*:}},
  ref={assumption~\theassumption.\Alph*}]

	\item Effective random assignment: $Z_i$ is independent of $D_i(z)$ and $Y_i(z)$, given  $\bm X$, for $z = 0,1$. More narrowly, we assume there are $n_1$ units randomly assigned to treatment according to complete randomization, leaving $n_0 = N-n_1$ units in control. We take $p =n_1/N$ with $p \in (0,1)$ as fixed.\footnote{For asymptotic arguments, this can be relaxed to just ensure that $n_1/N \to p$ and $N \to \infty$.} This implies  $Z$ is independent of $D_i(z)$ and $Y_i(z)$ unconditionally.
\item Exclusion restriction: $Z_i$ only impacts $Y_i$ through $D_i$, meaning there are no treatment impacts on anyone but the compliers, so $Y_i(1) = Y_i(0)$ if $D_i(1) = D_i(0)$.
	\item Relevance: $Z_i$ has a nonzero effect on $D_i$, i.e., we have at least some compliers in our dataset.
\end{enumerate} 
\end{assumption}
\noindent We note that under Assumption~\ref{assump:iv}, part A, the assumption of positivity: $0<P(Z=1) < 1$, is satisfied trivially.

Critically, under these assumptions, a causal quantity such as the average treatment effect is not point identified. There are two primary ways to achieve point identification. One way is to invoke some form of a homogeneity assumption and place a restriction on how the effects of $D_i$ and $Z_i$ vary from unit to unit in the study population. See \citet{hernan2019} and \citet{wang2018} for examples. Alternatively, one can invoke an assumption known as monotonicity:
\begin{assumption}
\label{assump:mono}
Monotonicity: $D_i(1) \geq D_i(0)$.
\end{assumption}
\noindent Monotonicity is best understood within a principal stratification framework \citep{Frangakis:2002}---but see \citet{AngImbRub96} for IV in particular. The principal stratification framework identifies that the members of our study population fall into distinct latent groups depending on their treatment exposure due to treatment assignment.
Under principal stratification unit $i$ is classified according to the following rules:
\begin{align*}
\text{complier} & \text{ if } D_i(1) = 1, D_i(0)=0,\\
\text{always-taker} & \text{ if } D_i(1) = 1, D_i(0)=1,\\
\text{never-taker} & \text{ if } D_i(1) = 0, D_i(0)=0,\\
\text{defier} & \text{ if } D_i(1) = 0, D_i(0)=1.
\end{align*}
Under the monotonicity assumption, we rule out the presence of ``defiers,'' or those who do the opposite of their assigned status.
The monotonicity assumption is integral to the methods we develop.

Next, let $\pi_c$ be the true proportion of compliers. Further let $n_c = \pi_cN$ be the number of compliers and $n_{c,z}$ be the number of compliers under treatment $z$. We use analogous notation for quantities related to always-takers and never-takers (e.g., $\pi_a$ and $\pi_n$ for the proportion of always-takes and never-takers, respectively). Define an indicator for being a complier as $C_i = 1$ if $D_i(1) = 1, D_i(0) = 0$, and $C_i = 0$ otherwise. Under monotonicity, we focus on the complier average causal effect (CACE) estimand:
\[ 
\CACE = \overline{Y}_c(1) - \overline{Y}_c(0) \mbox{ where } \overline{Y}_c(z) = \frac{1}{n_c}\sum_{i=1}^N C_i Y_i(z).
\]
One strength of the monotonicity assumption is that it does not impose any treatment effect homogeneity assumptions.
Specifically, the CACE only focuses on the effect among the complier population.
Under the exclusion restriction, the always-takers and never-takers have no measured impact as their treatment take-up does not change under random assignment.

Another useful causal quantity is the intention-to-treat effect. The ITT is $\overline{Y}(1) - \overline{Y}(0)$ with $\overline{Y}(z) = \frac{1}{N}\sum_{i=1}^N Y_i(z)$.
Due to the exclusion restriction and monotonicity we have
\[ 
\ITT = \frac{1}{N} \sum_{i=1}^N\left[ Y_i(1) - Y_i(0) \right]= \frac{1}{N} \sum_{i: C_i = 1} \left[Y_i(1) - Y_i(0) \right]+ \frac{1}{N} \sum_{i: C_i \neq 1} \left[Y_i(1) - Y_i(0)\right] = \pi_c \tau .
\] 

We treat both the CACE and ITT as finite-population quantities, meaning they are defined only with respect to the units in the experiment, the $n_c$ compliers and the $N$ units, respectively. We take potential outcomes to be fixed and, in the following section, estimators are random due solely to random assignment of units into treatment groups (as assumed given under Part A of Assumption~\ref{assump:iv}).
Thus expectations and variance are taken over the random treatment assignment process.

\subsection{IV Point and Variance Estimation}\label{sec:standard_iv_est}

The standard IV estimator for the CACE is
\[ 
\widehat{\CACE}_{\text{IV}} = \frac{ \widehat{\ITT} }{ \widehat{f} } ,
\]
where the ITT estimator of $\widehat{\ITT} = \overline{Y}^{obs}_1 - \overline{Y}^{obs}_0$, with
\[
\overline{Y}^{obs}_z = \frac{1}{n_z}\sum_{i: Z_i=z}Y_i(z), 
\]
is the numerator and the estimated proportion of compliers, $\pi_c$, is the denominator.
We can estimate $\pi_c$ via
\[
\widehat{f} =\overline{D}^{obs}_1 - \overline{D}^{obs}_0= \frac{1}{n_1}\sum_{i=1}^NZ_iD_i(1) - \frac{1}{n_0}\sum_{i=1}^N(1-Z_i)D_i(0). 
\]
Because $E[  \widehat{\ITT} ] = \pi_c \tau $ and $E[ \widehat{f} ] = \pi_c$, $\widehat{\CACE}_{\text{IV}}$ provides a reasonable estimate of $\tau$.
That being said, $\widehat{\CACE}_{\text{IV}}$, as a ratio estimator, will not be fully unbiased because $E[ A / B] \neq E[ A ]/ E[B]$ in general.

We can approximate the variance of $\widehat{\CACE}_{\text{IV}}$ using the delta method.
To define the finite-sample asymptotic variance, we require the following regularity assumptions:

\begin{assumption}\label{assump:IV_CLT}
Let $c(z) = \max_{1 \leq i \leq N} \left(Y_i(z) - \overline{Y}(z)\right)^2$. As $N \to \infty$,
\[\max_{z\in \{0,1\}}\frac{1}{n_z^2}\frac{c(z)}{n_0^{-1}S^2_Y(0) + n_1^{-1}S^2_Y(1) - N^{-1}S^2_Y(01)}\]
\end{assumption}

\begin{assumption}\label{assump:IV_CLT_uptake}
$\pi_{c}$, $\pi_{a}$, and $\pi_{n}$ have asymptotic limiting values such that at least two of those proportions are nonzero.\footnote{See Supplementary Material~\ref{app:var_cond_simple} for why this condition is sufficient for obtaining a central limit theorem result for treatment uptake. }
\end{assumption}

\begin{assumption}\label{assump:delta}
$N\text{var}(\widehat{\ITT})$ has a finite limiting value, to help ensure $\widehat{\ITT} - \ITT \overset{p}{\to} 0 $.
\end{assumption}
Next, we define the following variance expressions:
\[S^2_Y(z) = \frac{1}{N-1}\sum_{i=1}^N \left(Y_i(z) - \overline{Y}(z) \right)^2\]
and
\[S^2_Y(01) = \frac{1}{N-1}\sum_{i=1}^N \left(Y_i(1) -Y_i(0)  - \overline{Y}(1) + \overline{Y}(0) \right)^2.\]
\begin{align*}
S^2_D(1) &= \frac{N}{N-1}\pi_{n}(\pi_{c} + \pi_{a}),\\
S^2_D(0) &= \ \frac{N}{N-1}\pi_{a}(\pi_{c} + \pi_{n}),
\end{align*}
and
\begin{align*}
S^2_D(01) = \frac{N}{N-1}\pi_{c}(\pi_{a} + \pi_{n}).
\end{align*}

We can apply the finite-population Central Limit Theorem (CLT) framework to our $\widehat{\ITT} $ estimator under Assumption~\ref{assump:IV_CLT} based on Theorem 4 of \cite{LiDin17}. Similarly, we can obtain a finite-population CLT for treatment uptake, $\widehat{f}$, under Assumption~\ref{assump:IV_CLT_uptake}. 
By combining the CLT with Assumption~\ref{assump:delta} and an asymptotic  version of the relevance assumption (Assumption~\ref{assump:iv}.3), where $\pi_c$ has a nonzero limiting value, we can use the finite-population delta method to derive an asymptotic variance for $\widehat{\tau}_{\text{IV}}$ \citep{pashley2019note}. The formal expression for the asymptotic variance for $\widehat{\tau}_{\text{IV}}$ is
\begin{align}
\text{asyVar}\left(\widehat{\CACE}_{\text{IV}} \right) &=\frac{1}{\pi_c^2}\textrm{var}(\widehat{\ITT}) +\CACE^2\frac{1}{\pi_c^2}\textrm{var}(\widehat{f}) -2\CACE\frac{1}{\pi_c^2}\textrm{cov}(\widehat{\ITT}, \widehat{f}). \label{eq:delta_variance}
\end{align}
If all units are compliers, Equation~\ref{eq:delta_variance} collapses to $\textrm{var}(\widehat{\ITT})$.
We can rewrite the asymptotic variance as follows:
\begin{align*}
\text{asyVar}(\widehat{\CACE}_{\text{IV}})=&\frac{1}{\pi_c^2}\textrm{var}\left(\widehat{\ITT}- \CACE \widehat{f}\right) \\
=& \frac{1}{\pi_c^2}\frac{1}{(N-1)}\Bigg \{ \frac{1}{n_0}\sum_{i=1}^N\left( \tilde{Y}_i(0) - \overline{\tilde{Y}}(0) \right)^2 
  + \frac{1}{n_1}\sum_{i=1}^N \left(\tilde{Y}_i(1) - \overline{\tilde{Y}}(1) \right)^2 \\
& - \frac{1}{N}\sum_{i=1}^N \left(\widetilde{ITT}_i -  \widetilde{ITT}\right)^2  \Bigg \} ,
\end{align*}
where $\tilde{Y}_i(z) = Y_i(z) - \tau D_i(z)$ is an adjusted potential outcome based on $Y_i(z)$ and $D_i(z)$, so that $\widetilde{ITT}_i = Y_i(1) - Y_i(0) - \tau \left[D_i(1) -D_i(0) \right]$ and $\widetilde{ITT} = \tau\left[1-\pi_c\right]$.
In other words, the variance of our CACE estimate is equivalent to a $1/\pi_c^2$ scaling of the variance of an experiment where the average complier treatment impact has been subtracted off according to uptake behavior.
While the $\tilde{Y}_i(z)$ are always unobserved due to dependence on the estimand $\tau$, this formulation is important for further derivations detailed below and helps build intuition as to sources of variance for our CACE estimators.
Additional details on the asymptotic result are provided in Supplementary Material~\ref{append:iv_var}. 

Under the delta method, standard errors are estimated by plugging in estimates of all the terms in Equation~\ref{eq:delta_variance} \citep[see, e.g.,][chapter 23]{CausalInferenceText} as follows: 
\begin{align*}
\widehat{\text{var}}_{\text{DELTA}} \left(\widehat{\CACE}_{\text{IV}}\right) &= \frac{\widehat{\text{var}}\left(\widehat{\ITT}\right)}{\widehat{f}^2}+ \frac{\widehat{\ITT}^2\widehat{\text{var}}\left(\widehat{f}\right)}{\widehat{f}^4} - 2\frac{\widehat{\ITT}\widehat{\text{cov}}\left(\widehat{\ITT}, \widehat{f}\right)}{\widehat{f}^3}\\
&= \frac{1}{\widehat{f}^2}  \left( \frac{s_Y^2(1)}{n_1} + \frac{s_Y^2(0)}{n_0} \right)  
     + \frac{1}{\widehat{f}^4} \widehat{\ITT}^2\left(\frac{s_D^2(1)}{n_1} + \frac{s_D^2(0)}{n_0}\right) 
      - \frac{2}{\widehat{f}^3} \widehat{\ITT} \left(\frac{s_{Y,D}(1)}{n_1} + \frac{s_{Y,D}(0)}{n_0}\right)  
\end{align*}
where
\[s_{A,B}(z) = \frac{1}{n_z-1}\sum_{i:Z_i=z}\left(A_i(z) - \overline{A}^{obs}_z\right)\left(B_i(z) - \overline{B}^{obs}_z\right),
\]
and $s_A^2(z) = s_{A,A}(z)$, for $A= Y, D$ and $B= Y, D$.

One alternative method for estimating the IV standard error is the Bloom method (also sometimes called the Wald method), which essentially treats the denominator of our IV estimators as fixed \citep{Bloom:1984}. The Bloom method uses just the first term in the delta method expansion as the nominal variance of our estimator:
\begin{align*}
\text{var}\left(\widehat{\CACE}_{\text{IV}} \right) &\approx \frac{1}{\pi_c^2}\textrm{var}(\widehat{\ITT}).
\end{align*}
The plug-in estimate for this is then
\begin{align*}
\widehat{\text{var}}_{\text{BLOOM}}\left(\widehat{\CACE}_{\text{IV}}\right) 
    = \frac{\widehat{\text{var}}\left(\widehat{\ITT}\right)}{\widehat{f}^2}
    = \frac{1}{\widehat{f}^2} \left( \frac{s_Y^2(1)}{n_1} + \frac{s_Y^2(0)}{n_0} \right)  .
\end{align*}
The Bloom method will perform well if the second and third terms are small and/or cancel out. See \citet{Keele:2017fiv} for a detailed comparison of the delta and Bloom methods. 

\section{Post-stratification}

In many cases we might have a covariate that we believe to be predictive of complier status, or outcome, or both.
We would want to use this covariate to improve the precision of our IV estimator.
Motivated by the intuition that if we could isolate most of our compliers into a subset of our data then we could more reliably estimate impacts for those compliers as the instrument within that subset would be stronger, we turn to post stratification.

To implement post stratification, after randomization, we separate the units into $G$ groups based on some categorical (baseline) covariate that ideally predicts compliance type (or outcome). Because we only consider stratifying on covariates unaffected by treatment, group assignments are invariant to treatment assignment.
Note that the predictive potential of the covariates for compliance will not be used at an individual level to predict whether any single unit is a complier or to estimate a compliance probability function, but rather to form groups such that 
some groups have a high proportion of compliers and some have a lower proportion.
We assume the covariates used in the stratification are identified prior to looking at the sample used for estimating treatment effects.
If we instead search for covariates that are empirically predictive of compliance type (or outcomes) based on the observed data used for estimation, we run the risk of introducing additional bias into our estimation strategy; see, for example, \citet{Beach:1989wr} and \citet{Senn:1989wk} for the case of clinical trials.
We leave determining the practical consequences of violating this principle to future work.

We offer a few approaches to selecting covariates that are predictive of compliance type (or outcomes) to use for stratification.
First,  we could use data from a previous study to learn about relevant covariates.
For example, there is a rich literature of GOTV studies that can be used to learn about covariates that predict compliance type.

Second, if there is no relevant prior study available, one could rely on subject matter knowledge to choose the covariates for stratification.
For example, in our case knowing that people who are older are more likely to be home during the day and able to answer the door makes stratifying on age in studies with door-to-door canvassing potentially desirable.

Third, if sample size allows, we could use sample splitting to directly assess what is predictive of compliance.
In particular, we could split the sample and use one part as a training set to select relevant covariates and then perform the analysis with post-stratification the other part of the data based on those selected covariates.\footnote{It is possible to split the data in half and use this method on both halves of the data in an attempt to mitigate power loss, but to streamline theory we leave detailed investigation of such practice to future work.}

The theory in this section follows whether stratification is based on a covariate that is predictive or not.
In particular, stratifying on any baseline covariate will lead to valid inference, even if the covariate is useless.
However, as we will discuss in the next section, the \emph{benefits} of post-stratification will change based on the predictive power of the covariates.

Let $s_i = g$ if unit $i$ is assigned to group $g$. Let there be $N_g$ units in group $g$ with $N_{g,z}$ assigned to treatment $z$. We assume that $N_{g,z} \geq 1$ for $z\in\{0,1\}$ and $g = 1,\dots, G$.\footnote{In practice, the $N_{g,z}$ are random, and could be 0, depending on the treatment assignment. We discuss this further below.}
We can then apply each of the above estimators to each group $g$:
\begin{align*}
 \widehat{\ITT}_g= \frac{1}{N_{g,1}}\sum_{i:s_i=g}Z_iY_i(1) - \frac{1}{N_{g,0}}\sum_{i:s_i=g}(1-Z_i)Y_i(0),\\
 \widehat{f}_g = \frac{1}{N_{g,1}}\sum_{i:s_i=g}Z_iD_i(1) - \frac{1}{N_{g,0}}\sum_{i:s_i=g}(1-Z_i)D_i(0).
 \end{align*}

We initially offer two different estimators for a post-stratified CACE. First, we can post-stratify, followed by IV estimation within each stratum, with $\widehat{\tau}_g = \widehat{ITT}_g / \widehat{f}_g$. These $G$ IV estimators are then combined to estimate the overall CACE by weighting by the estimated number of compliers. Formally, this estimator is our ``IV-within'' estimator of
\begin{align}
\widehat{\CACE}_{\text{IV-w}} = \sum_{g=1}^G \frac{\widehat{f}_gN_g}{\sum_{k=1}^G\widehat{f}_k N_k} \frac{\widehat{\ITT}_g}{\widehat{f}_g}. \label{eq:IV-w}
\end{align}
If $\widehat{f}_g = 0$ then $\widehat{\tau}_g = \widehat{ITT}_g / \widehat{f}_g$ is undefined, but the weight $\widehat{f}_g N_g = 0$. 
We therefore define $\widehat{\CACE}_{\text{IV-w}}$ by dropping all strata where $\widehat{f}_g = 0$.
That is, we drop those portions of the experimental sample that we estimate as having no compliers from the analysis.
We also drop any strata $g$ that don't have at least 2 treatment and 2 control units (i.e., if $N_{g,z} \leq 1$ for $z = 0$ or 1), as we have no ability to estimate impact (if $N_{g,z}=0$) or standard errors (if $N_{g,z}=1$) otherwise.
The chance of $N_{g,z} \leq 1$ decays exponentially as sample size grows, making the bias of dropping such strata negligible. A stratum with 50 units in an RCT with treatment probability $0.3$ would have around a 1 in a million chance of being dropped, for example. See \citet{miratrix2013adjusting} for further discussion.

If $\hat{f}_g < 0$, then we have more always-takers in the treatment arm than control, and our $\hat{\tau}_g$ is based on a contrast of means not really related to the compliers, and, furthermore, the weight given to $\hat{\tau}_g$ is negative;
this motivates dropping all strata with 0 or non-negative weight. We explore these types of modifications further in Section~\ref{sec:weighting} and in the simulation study.

An alternate approach to post-stratification is to use the usual IV estimator, plugging in post-stratified estimates of the numerator term, $\widehat{\ITT}$, and denominator term, $\widehat{f}$.
This gives our ``IV-across'' estimator of:
\begin{align}
\widehat{\CACE}_{\text{IV-a}} & =  \frac{ \widehat{\ITT}_{\text{PS}}}{\widehat{f}_{\text{PS}}} =  \frac{ \sum_{g=1}^G\frac{N_g}{N}\widehat{\ITT}_g}{\sum_{g=1}^G\frac{N_g}{N}\widehat{f}_g}. \label{eq:IV-a}
\end{align}
We again drop strata with too few units to estimate standard errors or point estimates, as with $\text{IV-w}$.
Our first estimator, $\text{IV-w}$, calculates IV estimates \emph{within} the strata, and the second, $\text{IV-a}$, calculates one IV estimate \emph{across} the strata.
These estimators are related to each other and to a version of two-stage least squares, as the following two lemmas show.

First, Lemma~\ref{lemma:2sls_connection} outlines the equivalence between $\text{IV-a}$ and a two-stage weighted least squares estimation strategy.

\begin{lemma}\label{lemma:2sls_connection}
$\widehat{\CACE}_{\text{IV-a}}$ will equal $\hat{\beta}_{1, S2}$, the coefficient for predicted compliance in the second stage of a two-stage weighted least squares regression (with weighting in both stages) with weights $w_i = \frac{N_{g}}{N_{g,z}}\frac{n_z}{N}$ for unit $i$ in strata $g \in \{1,\dots,G\}$ assigned to treatment $z \in \{0,1\}$.
\end{lemma}
\noindent See Supplementary Material~\ref{append:2sls} for proof. 
With large strata, these weights are all approximately 1 given, for unit $i$ in group $g$, 
\begin{align*}
w_i = \frac{N_g}{N_{g,z}}\frac{n_z}{N} = \frac{1}{p_g} p \approx 1 ,	
\end{align*}
as random assignment will ensure each strata has roughly the same proportion treated, with $p_g \approx p$.
This suggests that in large samples with large strata, the weighted 2SLS estimate will generally be close to the usual, unweighted, 2SLS.
See \citet{schochet2024design} for an in depth look at regression adjusted instrumental variable estimators, which are shown to have similar asymptotic variances to our estimators if strata are not dropped.

Second, Lemma~\ref{lemma:w_a_connection} shows that $\widehat{IV}_w$ and $\widehat{IV}_a$ are equivalent when all strata have non-zero estimated proportions of compliers:

\begin{lemma}\label{lemma:w_a_connection}
If $\widehat{f}_g \neq 0$ for all $g = 1,\dots, G$, then the estimator in Equation~\ref{eq:IV-w} is mathematically equivalent to the estimator in Equation~\ref{eq:IV-a}:
\begin{align}
\widehat{\CACE}_{\text{IV-w}} &= \sum_{g=1}^G \frac{\widehat{f}_gN_g}{\sum_{k}\widehat{f}_k N_k} \frac{\widehat{\ITT}_g}{\widehat{f}_g} \nonumber\\
&= \sum_{g=1}^G  \left( \frac{N_g}{\sum_{k }N_k\widehat{f}_k} \right) \widehat{\ITT}_g \label{eq:weighted_ITT}\\
&=  \frac{ \sum_{g=1}^G\frac{N_g}{N}\widehat{\ITT}_g}{\sum_{k=1}^G\frac{N_k}{N}\widehat{f}_k} = \widehat{\CACE}_{\text{IV-a}} . \nonumber
\end{align}
\end{lemma}

The equivalence between $\widehat{\CACE}_{\text{IV-a}} $ and $\widehat{\CACE}_{\text{IV-w}}$ whenever $\hat{f}_g \neq 0$ for all $g=1,\dots,G$ allows us to use either representation to derive the asymptotic results.
Later, we find in simulations (with a fixed, finite sample) that $\widehat{\CACE}_{\text{IV-w}}$, which drops strata estimated to have zero compliers (as well as the variant that drops all non-positive estimates), has better properties than $\widehat{\CACE}_{\text{IV-a}}$.
This finding motivates estimators that weight those strata with more compliers more heavily as a means of achieving greater performance gains; see Section~\ref{sec:weighting}.

\subsection{Asymptotic Variance for Post-stratification Estimators}

Derivation of the asymptotic variance of $\widehat{\tau}_{IV-a}$ and $\widehat{\tau}_{IV-w}$ requires several finite-population central limit theorem results, which we include in the Supplemental Materials.
Let $W_i(g) = 1$ if $s_i = g$ and $W_i(g) = 0$ if $s_i \neq g$.
Then we can define variance components within each stratum:
\[S^2_{g,Y}(z) = \frac{1}{N_g-1}\sum_{i=1}^N W_i(g)\left(Y_i(z) - \overline{Y}_g(z) \right)^2,\]
\[S^2_{g,Y}(1,0)=  \frac{1}{N_g-1}\sum_{i=1}^NW_i(g)\left(Y_i(1) - \overline{Y}_g(1)\right)\left(Y_i(0) - \overline{Y}_g(0)\right),\]
and
\[S^2_{g, Y}(01) = \frac{1}{N_g-1}\sum_{i=1}^N W_i(g)\left(Y_i(1) - Y_i(0) -\left[ \overline{Y}_g(1) - \overline{Y}_g(0) \right]\right)^2.\]

Based on the conditions in the Supplemental Materials and using results from \cite{schochet2023design}, we get the asymptotic variance as given in the following theorem:

\begin{theorem}\label{thm:main_clt}
If we have Assumptions \ref{assump:iv}, \ref{assump:mono}, \ref{assump:strata_prop}, \ref{assump:li_ding_cond_main}, \ref{assump:clt_cond2_main}, \ref{assump:delta_post_strat}, and \ref{assump:clt_cond_d}, we have an asymptotic variance of
\begin{align*}
\text{asyVar}(\widehat{\CACE}_{\text{IV-a}}) =&\frac{1}{\pi_c^2}\text{asyVar}\left(\widehat{\ITT}_{\text{PS}}\right) +\CACE^2\frac{1}{\pi_c^2}\textrm{asyVar}(\widehat{f}_{\text{PS}}) -2\CACE\frac{1}{\pi_c^2}\textrm{asyCov}(\widehat{\ITT}_{\text{PS}}, \widehat{f}_{\text{PS}})
\end{align*}
where
\begin{align*}
\text{asyVar}\left(\widehat{\ITT}_{\text{PS}}\right) 
& =  \sum_{g=1}^N\frac{N_g}{N}\frac{N_g-1}{N-1}\left[\frac{S^2_{g,Y}(0)}{(1-p)N_g} +\frac{S^2_{g,Y}(1)}{pN_g} - \frac{S^2_{g, Y}(01)}{N_g}\right]\\
& \approx  \sum_{g=1}^N\frac{N_g^2}{N^2}\left[\frac{S^2_{g,Y}(0)}{(1-p)N_g} +\frac{S^2_{g,Y}(1)}{pN_g} - \frac{S^2_{g, Y}(01)}{N_g}\right]
\end{align*}
and 
\begin{align*}
\text{asyVar}(\widehat{f}_{\text{PS}}) 
&= \sum_{g=1}^G\frac{N_g}{N(N-1)}\left[(1-p)^{-1}\left(\pi_{g,c} + \pi_{g,a}\right)\pi_{g,n} + p^{-1}\left(\pi_{g,c} + \pi_{g,n}\right)\pi_{g,a} - \left(\pi_{g,a} + \pi_{g,n}\right)\pi_{g,c}\right]
\end{align*}
\label{thm:aymp_var_PS}
\end{theorem}
Theorem~\ref{thm:aymp_var_PS} comes from two finite-population central limit theorem results being combined with a finite-population delta method argument \citep{pashley2019note, schochet2023design}.
See Supplementary Material~\ref{append:clt_itt} for details along with finite-population CLT results for the ITT estimators. 

The asymptotic covariance term can similarly be approximated by a blocked covariance expression (see Supplementary Material~\ref{append:two_sided_bias}):
\begin{align*}
\textrm{asyCov}(\widehat{\ITT}_{\text{PS}}, \widehat{f}_{\text{PS}})
& \approx  \sum_{g=1}^N\frac{N_g^2}{N^2}\Bigg[\frac{\pi_{g,n}\pi_{g,c}}{p(N_g-1)}\left(\overline{Y}_{g,c}(1) - \overline{Y}_{g,n}(0)\right) + \frac{\pi_{g,n}\pi_{g,a}}{p(1-p)(N_g-1)}\left(\overline{Y}_{g,a}(1) - \overline{Y}_{g,n}(0)\right)\\
& \qquad \qquad \qquad + \frac{\pi_{g,a}\pi_{g,c}}{(1-p)(N_g-1)}\left(\overline{Y}_{g,a}(1) - \overline{Y}_{g,c}(0)\right) - \frac{\pi_{g,c}(1-\pi_{g,c})}{N_g-1}\CACE_g\Bigg],
\end{align*}
where $\overline{Y}_{g,t}(z)$ is the average potential outcome for units of compliance type $t \in \{a, c, n\}$ in stratum $g$ under treatment $z$.

Similar to the standard IV estimator, we can rewrite the above asymptotic variance expression for the stratified IV estimator as 
$$ \text{asyVar}(\widehat{\CACE}_{\text{IV-a}})  = \frac{1}{\pi_c^2}\textrm{var}\left(\widehat{\ITT}_{\text{PS}} - \CACE \widehat{f}_{\text{PS}}\right),$$
which corresponds to the post-stratified variance of a completely randomized experiment with potential outcomes $\tilde{Y}_i(z) = Y_i(z) - D_i(z)\tau$, where $\tau$ is the true CACE (across strata).

Due to the correspondence between $\widehat{\CACE}_{\text{IV-w}}$ and $\widehat{\CACE}_{\text{IV-a}}$, we further have $\text{asyVar}(\widehat{\CACE}_{\text{IV-w}}) = \text{asyVar}(\widehat{\CACE}_{\text{IV-a}})$ if we have constants $c_g >0$ such that $f_g \to c_g$ as $n \to \infty$ for all $g \in \{1,\dots,G\}$, guaranteeing that (asymptotically) IV-w does not drop any strata.

The Bloom approximation for the post-stratified estimator is 
\begin{align*}
\text{asyVar}\left(\widehat{\CACE}_{\text{IV-a}} \right) = \text{asyVar}\left(\widehat{\CACE}_{\text{IV-w}} \right) &\approx \frac{1}{\pi_c^2}\textrm{var}(\widehat{\ITT}_{\text{PS}}).
\end{align*}

\subsection{Estimating precision}

Standard errors for $\widehat{\CACE}_{\text{IV-w}}$ and $\widehat{\CACE}_{\text{IV-a}}$ are obtained by estimating variances within each stratum and aggregating.
For the Bloom method, for example, the across strata estimator is
\begin{align*}
\widehat{\text{var}}_{\text{BLOOM}}\left(\widehat{\CACE}_{\text{IV-a}} \right) &\approx \frac{1}{\widehat{f}_{\text{PS}}^2}\sum_{g=1}^G\frac{N_g^2}{N^2}\widehat{\textrm{var}}(\widehat{\ITT}_{g}).
\end{align*}
For each stratum $g$, we can estimate $\widehat{\textrm{var}}(\widehat{\ITT}_{g})$ if there are at least two units assigned to treatment and two units assigned to control.
If not, blocked variance estimators for blocks with only a single treated or control unit would need to be employed \citep{pashley2021insights}.

Although asymptotically equivalent, the delta method suggests different variance estimators for $\widehat{\CACE}_{\text{IV-a}}$ and $\widehat{\CACE}_{\text{IV-w}}$ where we drop strata with zero estimated compliers when estimating the variance for $\widehat{\CACE}_{\text{IV-w}}$.
If $N_{g,z} \geq 2$ for all $g=1,\dots,G$ and $z \in\{0,1\}$, the variance estimator for $\widehat{\CACE}_{\text{IV-a}}$ is:
\begin{align*}
&\widehat{\text{var}}_{\text{DELTA}} \left(\widehat{\CACE}_{\text{IV-a}}\right)\\
&=\frac{1}{\widehat{f}_{\text{PS}}^2} \sum_{g=1}^G\frac{N_g^2}{N^2}\Bigg(\frac{s_{Y,g}^2(1)}{N_{g,1}} + \frac{s_{Y,g}^2(0)}{N_{g,0}} + \left(\widehat{\CACE}_{\text{IV-a}}\right)^2\left(\frac{s_{D,g}^2(1)}{N_{g,1}} + \frac{s_{D,g}^2(0)}{N_{g,0}}\right) - 2\widehat{\CACE}_{\text{IV-a}}\left(\frac{s_{Y,D,g}(1)}{N_{g,1}} + \frac{s_{Y,D,g}(0)}{N_{g,0}}\right)\Bigg).
\end{align*}
\noindent This estimator is based on the modified potential outcome representation within strata, and summing across strata with weights $N_g^2/N^2$. A variance estimator for $\widehat{\CACE}_{\text{IV-w}}$ is the following weighted sum of stratum-level IV variance estimates:
\begin{align*}
&\widehat{\text{var}}_{\text{DELTA}}(\widehat{\CACE}_{\text{IV-w}})=\sum_{g=1}^G\frac{\hat{f}_g^2N_g^2}{N^2\widehat{f}_{\text{PS}}^2}\widehat{\textrm{var}}\left(\widehat{\CACE}_{\text{IV, g}}\right)\\
&= \sum_{g=1}^G\frac{\mathbb{I}(\hat{f}_g \neq 0)N_g^2}{N^2\widehat{f}_{\text{PS}}^2}\Bigg(\frac{s_{Y,g}^2(1)}{N_{g,1}} + \frac{s_{Y,g}^2(0)}{N_{g,0}} + \widehat{\CACE}_{\text{IV-w}}^2\left(\frac{s_{D,g}^2(1)}{N_{g,1}} + \frac{s_{D,g}^2(0)}{N_{g,0}}\right) - 2\widehat{\CACE}_{\text{IV-w}}\left(\frac{s_{Y,D,g}(1)}{N_{g,1}} + \frac{s_{Y,D,g}(0)}{N_{g,0}}\right)\Bigg),
\end{align*}
where $\mathbb{I}(\hat{f}_g \neq 0) = 1$ if $\hat{f}_g \neq 0$ and is 0 otherwise.
It is straightforward to see that $\widehat{\text{var}}_{\text{DELTA}}(\widehat{\CACE}_{\text{IV-a}}) = \widehat{\text{var}}_{\text{DELTA}}(\widehat{\CACE}_{\text{IV-w}})$ whenever all $\hat{f}_g$ are nonzero.
Again, if there are not enough units per stratum to estimate variance within each, a combined blocked variance estimator would be necessary \citep{pashley2021insights}.

\section{Benefits of Post-stratification for IV estimates}

In this section we analytically derive three potential benefits from post-stratification for IV estimators. Specifically, we focus on when post-stratification (1) reduces variance, (2) increases precision of the estimated standard errors, and (3) reduces bias. 

\subsection{Variance Reduction and Standard Errors}

Post-stratification on a covariate predictive of either complier status or the outcome would ideally reduce the variance of our IV estimator. 
However, covariates predictive of being a complier and those predictive of outcome are not equal in terms of their ability to increase precision, especially if we are not dropping low-complier strata from the analysis.

Compare the delta-method asymptotic variances for the IV estimator before and after post-stratification:
\begin{align*}
\text{asyVar}(\widehat{\CACE}_{\text{IV}}) &= \frac{1}{\pi_c^2}\textrm{var}\left(\widehat{\ITT}- \frac{\ITT}{\pi_c}\widehat{f}\right) \mbox{ and } \\
\text{asyVar}(\widehat{\CACE}_{\text{IV-a}}) &= \frac{1}{\pi_c^2}\textrm{var}\left(\widehat{\ITT}_{\text{PS}} - \frac{\ITT}{\pi_c}\widehat{f}_{\text{PS}}\right) .
\end{align*}
As shown above, we can view both these asymptotic variances as exactly the (scaled) variances we would get from running a completely randomized experiment with potential outcomes $\tilde{Y}_i(z) = Y_i(z) - \frac{\ITT}{\pi_c}D_i(z)$, either unadjusted or with post-stratification, respectively.

The comparison of asymptotic variance for $\widehat{\CACE}_{\text{IV}}$ vs $\widehat{\CACE}_{\text{IV-a}}$ therefore amounts to whether post-stratification would be beneficial in an experiment with potential outcomes $\tilde{Y}_i(z) = Y_i(z) - \frac{\ITT}{\pi_c}D_i(z)$.
Based on \citet{miratrix2013adjusting}, we should expect (informally) for post-stratification to be beneficial in terms of variance reduction the more the variability of $\tilde{Y}_i(1)$ and $\tilde{Y}_i(0)$ is between strata than within.
From this result, to reduce variability through post-stratification we might consider reducing within stratum variability via either of the two pieces of $\tilde{Y}_i(z)$: make units within each stratum similar in terms of potential outcomes, $Y_i(z)$, without regard to the compliance aspect, or make units within each stratum similar in terms of $D_i(z)$ (compliance type).\footnote{A third option would be to target the entire expression $Y_i(z) - \frac{\ITT}{\pi_c}D_i(z)$, but we believe that it would be difficult to find covariates to do this directly.}
We expect reducing overall variation by targeting variation in compliance to be difficult for at least two reasons.
First, because the $D_i(z)$ terms are multiplied by the CACE in the $\tilde{Y}_i(z)$ expressions, if the CACE is small relative to overall variation in $Y_i(z)$, the impact of targeting $D_i(z)$ will likely be minor.
Second, if compliance is relatively rare, then the number of units that are actually differentially adjusted across treatment arms will be few, again making the impact of the second term minor.
Put differently, if the CACE is 0, then even if we stratify perfectly on compliance type, we will only have gains if this stratification were effective for the original $Y_i(z)$, meaning our stratification variable was predictive of the original $Y$s as well.

Regarding the precision of standard errors, we can again use our post-stratification results for the modified potential outcomes.
In particular, if we reduce the true variance of our estimator, we should also expect to reduce the variance in our estimate of that variance \citep{pashley2020block}.
In our simulation study, we explore the extent of improvement in the standard error estimates.

Note that we focused here on the comparison of $\widehat{\CACE}_{\text{IV}}$ with $\widehat{\CACE}_{\text{IV-a}}$ rather than $\widehat{\CACE}_{\text{IV-w}}$.
Recall  $\widehat{\CACE}_{\text{IV-a}}$ and $\widehat{\CACE}_{\text{IV-w}}$ are the same when there are no strata with zero estimated compliers.
Our simulations in Section~\ref{sec:sims} demonstrate that the feature of $\widehat{\CACE}_{\text{IV-w}}$ of dropping strata with no estimated compliers reduces variability, and that these benefits can be even greater when we drop or down-weight strata more aggressively, as we will discuss in Section~\ref{sec:weighting}.

\subsection{Bias}

Post-stratification can also reduce the bias in the IV estimates that comes from the ratio estimator.
Critically, the bias reduction depends on whether noncompliance is one or two-sided.
That is, in some applications controls are unable to access treatment receipt such that $P(D_i = 0|Z_i=0) = 1$.
This is referred to as one-sided noncompliance. When this does not hold, there is two-sided noncompliance. 

In our calculations below, bias is with respect to the finite-population CACE under Assumptions~\ref{assump:iv} and \ref{assump:mono}.
The bias exists even under these assumptions (including when the IV is randomized) due to the random denominator in the IV estimator.
If the desired target of inference is the average effect for the full sample of $N$ units (assuming some hypothetical intervention in which noncompliers could be forced to comply) or a larger population from which the units were sampled, there would be additional generalizability bias.
For estimators that potentially drop or down-weight strata (e.g., $\widehat{\CACE}_{\text{IV-w}}$), we can have within-sample generalizability bias if the compliers in the dropped strata have systematically different treatment effects than the compliers in the kept strata; we discuss this in detail when we extend our family of estimators further in Section~\ref{sec:weighting}.

We first more precisely characterize the possible bias reduction due to post-stratification with one-sided noncompliance.
Under one-sided noncompliance, $\hat{f}$ is the observed proportion of those who took treatment in the treatment group.
When noncompliance is one-sided, we can write the bias in the standard IV estimator as:
\begin{align}
E\left[\widehat{\CACE}_{\text{IV}}\right] - \CACE 
&=  \frac{1}{1-p}\left(1  - E\left[ \frac{1}{\hat{f}}\right]E[\hat{f}]\right)\left( \overline{Y}_c(0)-\overline{Y}_n(0)\right) = \frac{1}{1-p}\text{cov}\left(\hat{f},  \frac{1}{\hat{f}}\right)\left( \overline{Y}_c(0)-\overline{Y}_n(0)\right) . \label{eq:bias}
\end{align}
\noindent The derivation of this result can be found in Supplementary Material~\ref{append:one_sided_bias}.

The dependence of Equation~\ref{eq:bias} on the covariance between $\hat{f}$ and $1/\hat{f}$ and a further Taylor approximation given in Supplementary Material~\ref{append:one_sided_bias} reveals that the bias depends on the magnitude of the $\text{var}(\hat{f})$. 
As such, reducing the variance of $\hat{f}$ can also reduce bias in the estimator. Therefore, the variance reduction properties of post-stratification can reduce bias as well. We can also characterize the direction of the bias reduction.
The bias reduction depends on the relative averages of outcomes of the compliers under control and the never-takers (under control or treatment).
We can express this quantity as $\Delta = \overline{Y}_c(0)-\overline{Y}_n(0)$, where $\overline{Y}_c(z)$ is the average potential outcome for all compliers in the sample under treatment $z$, and $\overline{Y}_a(z)$ and $\overline{Y}_n(z)$ are averages for the always- and never-takers.
A negative $\Delta$ implies the bias will be positive, and a positive $\Delta$ implies the bias will be negative.

When noncompliance is two-sided, we can express the bias in the standard IV estimator as
\begin{align*}
E\left[\frac{\widehat{\ITT}}{\hat{f}}\right] - \CACE & \approx  \frac{1}{\pi_c^2} \left[ \CACE \text{var}(\hat{f})- \textrm{cov}(\widehat{\ITT}, \hat{f}) \right] \\
& = \frac{1}{\pi_c^2(N-1)}\Bigg[ \frac{\pi_n((1-p)\pi_c + \pi_a)}{p(1-p)}(\overline{Y}_n(0) - \overline{Y}_c(0)) + \frac{\pi_a(p\pi_c + \pi_n)}{p(1-p)}(\overline{Y}_c(1) -\overline{Y}_a(1) )\Bigg]
\end{align*}
\noindent See Supplementary Material~\ref{append:two_sided_bias} for the derivation. In this context, the bias again depends upon the magnitude of the variance of $\hat{f}$. However, characterizing the direction of the bias is more complicated than in the one-sided noncompliance case.
The direction of the bias now depends on the relative group means of compliers, always-takers, and never-takers.
The relative differences between these three groups can either offset or increase the bias terms. Specifically, there will be a positive bias if $\overline{Y}_n(0) > \overline{Y}_c(0)$ and $\overline{Y}_c(1) >\overline{Y}_a(1)$, and there will be a negative bias if $\overline{Y}_n(0) < \overline{Y}_c(0)$ and $\overline{Y}_c(1) <\overline{Y}_a(1)$.

In this section, we focused on the possibility of bias reduction due to reducing variability.
However, the story is more complicated if, as we do with $IV_w$, we drop or re-weight strata, which can \emph{introduce} bias.
In particular, if some strata have higher chances of being dropped or down-weighted than others, then this can cause bias with respect to the overall CACE estimand.
See Section~\ref{sec:weighting} for further discussion.
Additionally, if our assumptions do not hold, additional bias can enter through post-stratification.
In particular, we show in Supplementary Material~\ref{supsubsec:er} that post-stratification can amplify bias if the exclusion restriction does not hold.

\section{Alternative post-stratification strategies}
\label{sec:weighting}

The original intuition of our post-stratification approach was that if we could isolate compliers into a few strata, we could benefit by the improved estimation in those strata.
The story turns out to be more complex than this, in that strata with few compliers are so unstable that, even though they have less overall weight in the final estimate, they are so variable that they can undo the precision gains achieved by having a greater proportion of compliers in the high-complier-rate strata.

To see this, consider the second line of Equation~\ref{eq:weighted_ITT} (Lemma~\ref{lemma:w_a_connection}): this line shows $IV_a$ (and thus $IV_w$ with no strata dropped) as a weighted average of ITT estimates, with the weights of the strata not dependent on the number of compliers, but instead the overall strata sizes.
This weighting by strata size means our strategy to upweight complier-rich strata is ineffective. 
In particular, even if stratum $g$ has few compliers, it will contribute just the same to the overall CACE estimate as it would to an ITT estimate!

If we are willing to possibly incur some further bias (beyond the normal bias of an IV estimator), we could decrease the weight of those strata with fewer compliers to potentially achieve precision gains by avoiding the weak instrument problem.
In this section we discuss two strategies for achieving this that build upon post-stratification.
The first strategy is to outright prune those strata with few compliers, similar to trimming an observational study of hard-to-match units.
The second strategy is to weight strata roughly proportional to complier prevalence or CACE estimator precision.

The $IV_w$ estimator already drops strata when $\widehat{f}_g = 0$ and, as we will see in the simulation study, even this limited pruning can stabilize the overall estimator.
If, however, $\widehat{f}_g \approx 0$ then we would include the stratum estimate of $\widehat{\ITT}_g/\widehat{f}_g$, which will be quite unstable due to the small denominator, in the overall weighted average.
We can avoid including such unstable estimates by thresholding at something other than exactly 0.
For example, we could drop any strata with estimated proportion of compliers less than 2\% (or any threshold of our choosing); we call this estimator the ``Drop-Small-Strata'' (DSS) estimator.
An alternative version of this estimator, which we call the ``Drop-Small-F'' (DSF) estimator, drops any stratum from the estimator where we fail the weak instrument test \citep{Stock:2005} for that stratum based on the F statistic for the importance of $Z$ in predicting $D$.
In particular, following common practice for IV estimation, we drop those strata where $F < 10$.

Dropping strata with low proportion of compliers has two benefits: first, we avoid unstable and potentially very large estimates from the low-compliance strata.
In other words, we focus our weighted average of ITT estimates on those strata with more compliers, and thus on those with more information on the treatment effect of compliers.
Second, we avoid odd behavior in the two-sided noncompliance case when $\widehat{f}_g < 0$.
For a stratum with $\widehat{f}_g < 0$, the estimate of the ITT is a function of the overabundance of always-takers in the control side, and a sign flip due to the negative compliance rate, neither of which have anything to do with treatment impact for the compliers in the stratum.
Given this observation, we introduce the Drop-Small-Strata-less-than-0 (DSS0) estimator, which just drops all strata with $\hat{f}_g \leq 0$.

For an alternate approach, consider that post-stratification is a weighted average of subgroup estimates. 
We saw that we weighted the CACE estimates for $\widehat{\CACE}_{\text{IV-w}}$ by the estimated number of compliers within each stratum, and the component parts of $\widehat{\CACE}_{\text{IV-a}}$ by the number of individuals within each stratum.
We might naturally wonder about other weightings of these component parts.
In particular, we might weight by the (estimated) precision of each strata using a Precision Weighted IV estimator (PWIV).
The PWIV estimator has a form similar to $\widehat{\CACE}_{\text{IV-w}}$ but weights by the (Bloom) estimated precision of each stratum: 
\[
\widehat{\tau}_{\text{PWIV}} = \frac{1}{W}\sum_{g=1}^G \frac{\widehat{f}_g^2}{\widehat{\text{var}}(\widehat{\ITT}_g)} \widehat{\CACE}_{g} \mbox{ with } W = \sum_{h=1}^G \frac{\widehat{f}_h^2}{\widehat{\text{var}}(\widehat{\ITT}_h)}.
\]
This estimator more heavily weights strata with higher estimated proportion of compliers than $IV_w$ (see the $\hat{f}^2_g$ term in the weight, vs. $\hat{f}_g$ in $IV_w$).

\noindent \textbf{Impact of treatment effect heterogeneity}

 Either by dropping strata or reweighting them, the overall goal is to discount those strata with low proportions of compliers, as their associated estimates are very unstable, and focus attention on strata where, in principle, estimating the complier average impact is easier.
This can incur bias: this process will shift our estimand away from the overall CACE, and towards a reweighted CACE tilted towards the average effect of those compliers in strata that tend to be kept or upweighted.
If these compliers have, for example, higher impacts in general, our overall estimates will tend to be larger than the true overall CACE.
That being said, if the CACE is homogenous across strata, or if the proportion of compliers left out of the overall estimate is small, than this bias will be minimal.
Especially considering we are targeting dropping strata with few compliers, it seems reasonable that our biased CACE may not be too far off the true overall CACE target.
In other words, the more homogenous the treatment effect is across all units (in the sample or population that is the target of inference), the smaller any generalizability bias.
As such, a useful diagnostic would be to consider the possibility of  treatment effect heterogeneity among the compliers. In general, this will consist of qualitative arguments, since the compliers are unobserved. 
That being said, it may be difficult to reason about treatment effect heterogeneity among the compliers when multiple categorical variables are combined to create relatively smaller strata.

As an additional problem, we are dropping or down-weighting strata based on \emph{estimated} compliance rates.
 This can create difficulties; for example, under one-sided noncompliance, we only drop those strata with no compliers randomized to the treatment assignment arm, as that is where we estimate the compliance rate, but keep strata if there are no compliers on the control side.
This asymmetry in our estimation approach again opens the door to bias.
The practical implications of all of these biases are explored further in the simulation section.

As a further advantage of these estimators, focusing attention on the complier-rich strata could partially control bias in contexts where the exclusion restriction is violated.
In particular, the bias caused by violation of the exclusion restriction via treatment impact on the noncompliers would generally be attenuated in the strata with a higher proportion of compliers, as there are fewer noncompliers represented in the ITT estimate.
The bias is, in other words, isolated into the low-compliance strata, which are then discounted in the overall estimate.
This is illustrated in the final additional simulation in the supplementary materials.
We note, however, that such attenuation of bias is not guaranteed through post-stratification.
For example, if noncompliers with the largest violation of the exclusion restriction, in the sense of having the largest effect of assignment of treatment on outcome, are precisely those noncompliers who end up in strata with compliers and therefore are not dropped, it's possible to see an increase in bias.

\section{Simulation Study}
\label{sec:sims}

We explore the analytic results using a simulation study to compare the different estimation strategies.\footnote{R package and replication files for simulations and GOTV applications can be found at \url{https://github.com/lmiratrix/poststratIV}.}
We investigate $\widehat{\CACE}_{\text{IV-a}}$ and $\widehat{\CACE}_{\text{IV-w}}$, the two stratification approaches described above.
Of course, as we have mathematically shown, these estimators are identical except when some strata have $\hat{f}_g = 0$; we will examine which estimator tends to perform better overall, including when this event occurs.
We also include the variant of $\widehat{\CACE}_{\text{IV-w}}$ where we drop all strata with $\hat{f}_g \leq 0$, the Drop-Small-F (DSF), the Drop-Small-Strata-less-than-0 (DSS0), and the Precision Weighted IV (PWIV) estimators from Section~\ref{sec:weighting}.

As a baseline we consider the simple IV estimator that ignores the covariate entirely. We also include the usual two-stage least squares estimator \citep[as implemented by the AER package in R,][]{aer}, using the stratification category as a covariate. We finally include an Oracle estimator of the simple difference in means estimate applied to the subset of compliers; this represents a best-case context where we know complier status perfectly.

For each estimator (save the Oracle and 2SLS) we have two methods for calculating a standard error: (1) the Bloom approach, where we consider the proportion of compliers as fixed, and (2) the delta method approach, which accounts for the uncertainty in estimating the compliance rate.
We compare the performance of these two standard error estimators along with the performance of the point estimators.

For each simulated dataset, we generate four strata, in line with our empirical example.
We vary several simulation factors of interest:

\begin{enumerate}
	\item Overall size of the experiment ($N = 500, 1000, 2000$).
	\item Overall proportion of compliers (5\%, 7.5\%, and 10\%).
	\item One-sided noncompliance, and two-sided noncompliance with two-thirds of the noncompliers being always-takers.
	\item Whether group membership predicts complier status or not.
	\item Whether group membership predicts the outcome or not.
	\item Whether the mean of the never-takers is below, equal to, or above the mean of the compliers' control potential outcomes. (We leave the always-takers mean in line with the compliers.)
	\item Whether the treatment impact is different across strata, or constant.
\end{enumerate}

The above factors result in 432 distinct scenarios that we explore.
For each iteration of our simulation, we generate all the potential outcomes for all units, and then randomize 30\% of the units into treatment, leaving the rest as controls.
We then apply our suite of estimators to the resulting data, and record the point estimate along with the two standard error estimates from the Bloom and Delta method approaches.
See Supplementary Materials~\ref{app:simulation} for further details on the data generating process.

In running our simulation, given the low compliance rates, some of the estimators could give extreme values, especially in the two-sided noncompliance case.
For example, the baseline unstratified estimator is undefined if the compliance rate is estimated at precisely 0, and can be of very large magnitude if the difference in treatment take-up in the two arms is a single unit.
The DSF estimator would often drop all strata in many of the two-sided noncompliance cases.
The 2SLS estimator similarly evidenced unstable behavior.
We therefore dropped all undefined trials and windsorized all impact estimates to $\pm 10$ standard deviations.
Overall, we dropped less than a tenth of a percent of our estimates (other than the DSF which was near 31\%), and windsorized a bit over 1\% of the remaining observations for 2SLS, $IV_a$, and the baseline, around 0.75\% of $IV_w$, and less for the remaining estimators.

The standard errors are also susceptible to low complier estimates, and can be too large.
In the two-sided noncompliance case, in particular, the asymmetrical treatment assignment means the estimated proportion of compliers can be very close to zero without being exactly zero, creating serious instabilities in the delta-method standard errors.\footnote{For example, consider a strata with 231 units, 76 treated and 155 control. Of the 76, 51 take treatment, and of the 155, 104 take treatment.  The estimated complier rate is then $51/76 - 104/155 \approx 8.5 \times 10^-5$, making the CACE estimate 11,780 times the ITT estimate for that strata, which is massive and implausible.  This extreme estimate would then get averaged, weighted by strata size, ruining the overall estimate. The unstratified estimator can also have this behavior.}
We thus windsorized the standard errors to $\pm 10$ standard deviations as well. Underscoring the instability in uncertainty estimation, around 5\% of the two-sided simulation trials were thus windsorized for each estimator other than DSF and PWIV.

\subsection{Results}

\begin{figure}[hbt]
  \includegraphics[width=\textwidth]{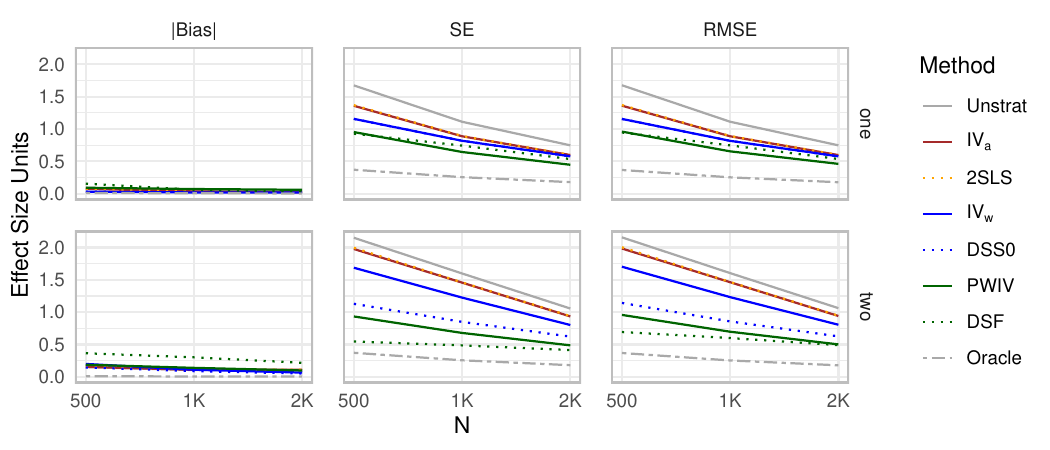}
  \caption{Overall performance characteristics of the point estimators with absolute bias, true Standard Error, and RMSE averaged across the different simulation scenarios grouped by overall sample size. Top row is one-sided noncompliance, bottom row is two-sided. Legend orders estimators by overall height of the lines. In particular, all considered estimators lie between the unstratified (top line) and oracle (bottom line) in performance.}
  \label{fig:overall_performance}
\end{figure}

Overall performance characteristics across all simulation scenarios are shown in Figure~\ref{fig:overall_performance}, with the top row being one-sided noncompliance and the bottom row being two-sided noncompliance.
We average the performance metrics across all the scenarios of a given sample size to get average trends across the other specifications.
In general, the post-stratified estimators have less bias, lower standard errors (SEs), and lower root mean square errors (RMSEs) than doing a standard unstratified analysis.
The 2SLS and $IV_a$ estimators basically coincide, as anticipated, in terms of performance; the lines are over-plotted in the figure.
The $IV_w$ estimator substantially outperforms the $IV_a$ estimator on average, and the weighting estimators are in turn outperforming $IV_w$ in terms of precision and overall RMSE.
The DSS0 estimator further outperforms $IV_w$ indicating that it is sensible to drop strata with zero or negative estimated proportion of compliers (which would indicate zero or negligible compliers under monotonicity), although the two estimators will always coincide in the case of one-sided noncompliance.
Relative to the standard error, the bias is negligible, although there is notably more bias for the two-sided non-compliance scenarios.
Overall, two-sided non-compliance is a harder estimation problem; bias, SE, and RMSE are all notably higher.

We verified that our two versions of post-stratification, $IV_a$ and $IV_w$, give identical point estimates if all strata are defined, as anticipated.
Notably, dropping strata estimated to be have zero compliers does impact \emph{overall} performance, allowing $IV_w$ to improve over $IV_a$.
Note how, in Figure~\ref{fig:overall_performance}, the average standard error and RMSE of $IV_{a}$ (IV across, using adjusted numerator and denominator for the overall ratio) is larger than that for $IV_{w}$ (IV within and then average).
We calculated the ratio of the RMSEs of $IV_{w}$ vs. $IV_{a}$ for all scenarios, and found that $IV_w$ can easily be more than 30\% smaller (with a 13\% average reduction across all scenarios) with the gains being correlated to the chance of dropped strata.
As a point of reference, across the simulation scenarios, one or more strata had zero estimated compliers (and were thus dropped) about half of the time due to a mix of small strata, small sample size, and low overall proportion of compliers.

\subsubsection{Variance reduction}

\begin{figure}[hbt]
  \includegraphics[width=\textwidth]{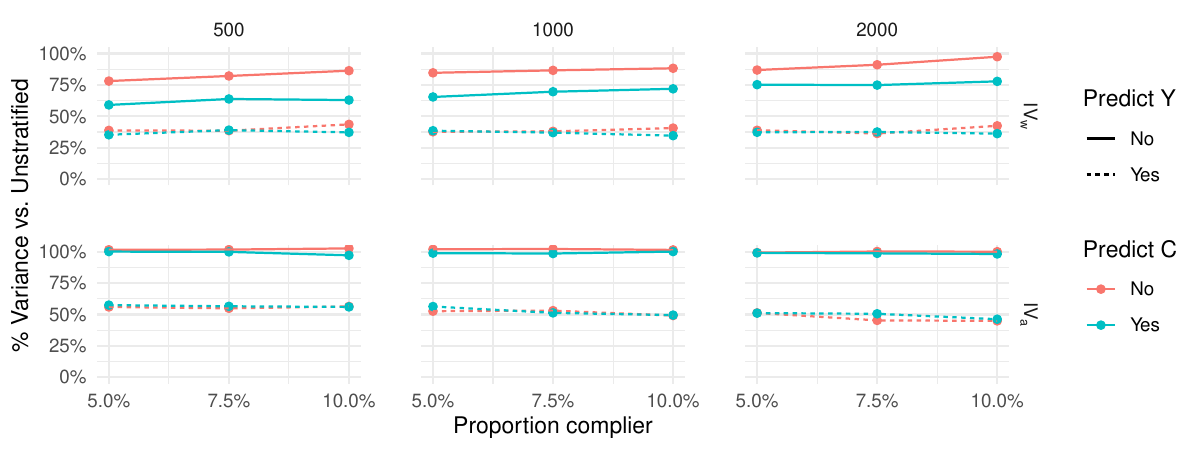}
  \caption{Ratio of variance of $IV_{w}$ to unstratified IV.}
  \label{fig:variance_ratio_plot}
\end{figure}

We next examine which factors drive the degree of improvement in the standard errors.
To contextualize the uncertainty, we compare the standard error of $IV_a$ and $IV_w$ to the unstratified standard error for each simulation scenario.
A ratio of 75\% for $IV_{w}$, for example, would represent a 25\% reduction in the variance if one post-stratifies using $IV_{w}$ vs. using the unstratified IV.
Figure~\ref{fig:variance_ratio_plot} shows that stratifying by a covariate predictive of $Y$ (outcome) or $C$ (compliance status) both help.
It is clear that using covariates predictive of outcome can substantially improve precision.
For covariates predictive of being a complier, gains are largest in the cases of low compliance and smaller sample sizes.
This is driven by dropping strata with 0 estimated compliers; see the second row for $IV_a$ that show no real gains of compliance-predictive covariates for $IV_a$ (and thus, by extension 2SLS).

For $IV_w$, we actually see a benefit even when the covariate is neither predictive of compliance status nor outcome, when $\pi_c$ and sample size is low (see top left of figure).
This stems from the benefits of dropping those strata with no observed compliers, which adds substantial stability to the estimator, providing benefits well beyond the bias incurred.
We unpack this surprising finding in the supplementary materials.
It is related to the discussion of why a complier-predictive covariate fails to provide much gains for 2SLS, which we discuss further in Section~\ref{sec:complier_predictive_covariate}

\subsubsection{Bias}

\begin{figure}[hbt]
  \includegraphics[width=\textwidth]{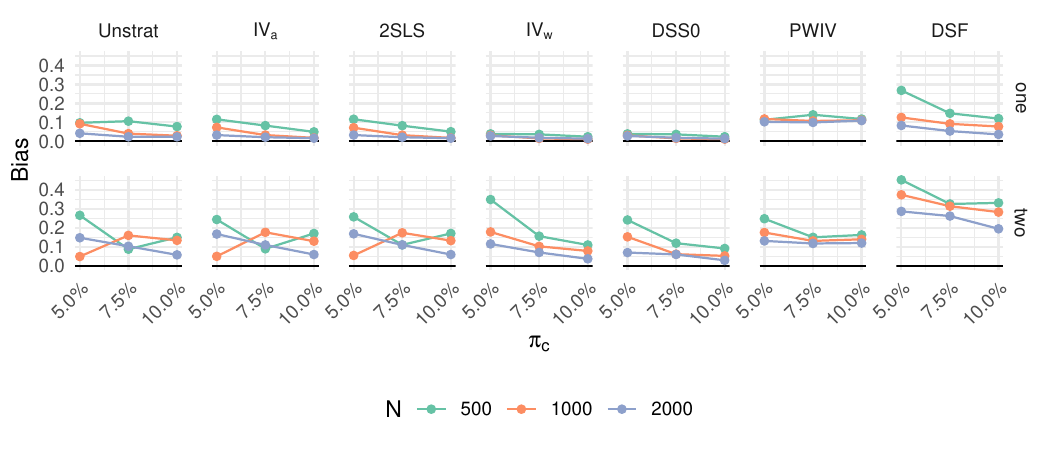}
\caption{Bias of estimators for heterogenous treatment effect scenarios only. Scenarios grouped by overall compliance rate ($x$-axis), sample size (color of line) and one- (top row) vs two-sided (bottom row) noncompliance.}
  \label{fig:bias}
\end{figure}

For the scenarios considered, bias is a much smaller share of the overall RMSE than variance; see the low range of biases on Figure~\ref{fig:overall_performance}, relative to the SEs.
We plot the biases again on Figure~\ref{fig:bias}, to clarify the bias trends.
For one-sided noncompliance, $IV_w$ has much less bias, as anticipated.  The instability it faces in the two-sided noncompliance case seems to erase bias gains, however (see bottom row).
As expected, DSF and PWIV have elevated levels of bias: they are estimating the CACE of the kept strata in the first case, and weighting the higher-compliance strata more heavily in the second case, which shifts their respective estimands. 
$IV_w$ and DSS0, in principle, do the same by dropping low-complier strata when the estimated proportion of compliers is precisely 0 (or negative).
Even so, the bias reduction due to stratification for this estimator cause it to generally be the least biased of all the estimators considered.
$IV_a$ (and by extension 2SLS) can have less bias than the simple IV, but the relative improvement is negligible across scenarios.

\subsubsection{Standard error estimation}

A standard error estimator is well calibrated if it is, on average, equal to the true standard error in a given context.
To assess this we calculate the square root of the ratio of the average of the squared standard error estimates to the true squared standard error (the estimator variance) for each context.\footnote{We calculate the ratio of variances because usual standard error estimators generally give unbiased variance, not SE, estimates. The outer square root brings the ratio back to the scale of standard error.}
Unfortunately, the standard error estimates, especially from the Delta method, have a very strong right skew, with fairly common extreme values.
We windsorized at a fairly large 10 standard deviations, but even so around 20\% of the standard errors were windsorized for some scenarios for the primary estimators.
This process will substantially reduce the average estimated standard error, which will play a role in interpreting the ratio of average to estimated true standard error.

\begin{figure}[hbt]
\center
  \includegraphics{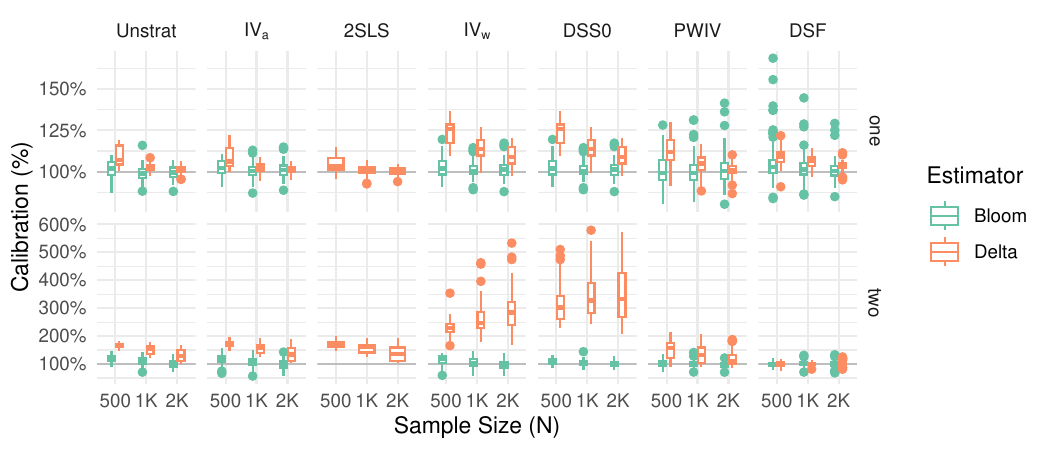}
  \caption{Relative percent change of average estimated standard error to true standard error, calculated as the square root of the mean estimated squared standard error divided by true variance (as estimated across simulation trials) for both the delta method and Bloom standard errors. Each point is a specific simulation scenario. Points above 100\% indicate standard errors systematically too large, and below systematically too small.}
  \label{fig:se_estimator_plot}
\end{figure}

Calibration results are on Figure~\ref{fig:se_estimator_plot}.
For one-sided noncompliance, the delta method for calculating standard errors can give standard errors that are a bit too high (15\% or more for many scenarios when $N=500$, for example), while the simpler Bloom estimator generally performs well, although they can be anti-conservative when sample sizes are small.
The 2SLS standard errors are also somewhat inflated for small sample size.
The DSF estimator, and to a lesser extent the PWIV estimator, tend to have overly large Bloom standard errors for some scenarios.
Two-sided noncompliance is a much worse story---note the $y$-axis has different scales in the top and bottom row of the figure---even with windsorizing, the average standard error regularly being 3 or 4 times too large.

Figure~\ref{fig:se_estimator_plot} is driven by the outliers; the ratio of the median estimated standard error to the true standard error (see supplement) is generally slightly below 100\%, indicating that, more often than not, the estimated standard error is too low. 
In Supplementary Material~\ref{sec:se_stability} we show that, for one-sided noncompliance, the stability of the estimated standard errors for the post-stratified estimators, relative to their true standard errors, is about the same as for the unstratified estimators.

Overall, these results underscore the difficulty of estimating uncertainty in weak instrument contexts.
That being said, the Bloom estimator does appear to be less vulnerable to extreme estimates and has decent overall properties.

\subsection{Complier Predictive Covariates}
\label{sec:complier_predictive_covariate}

\begin{figure}[hbt]
\centering
  \includegraphics[width=\textwidth]{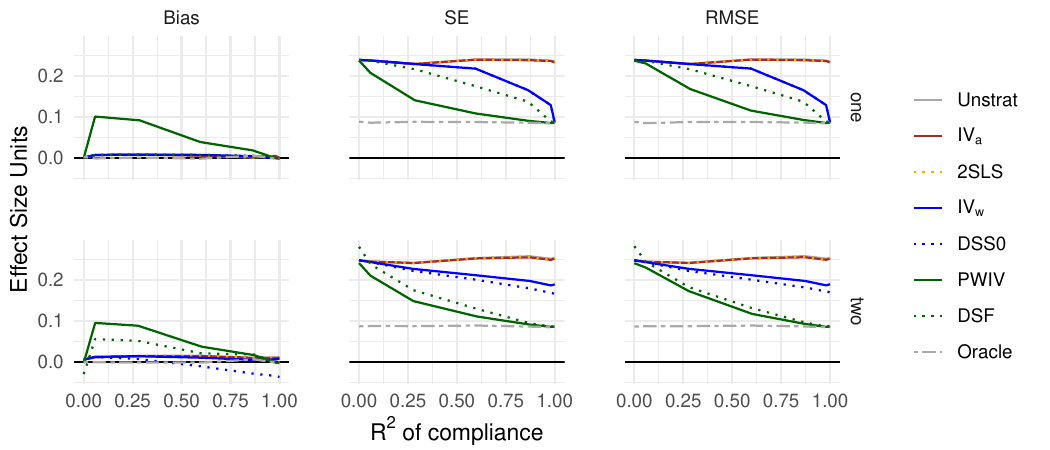}
 \caption{Simulation results with compliers increasingly concentrated in the upper strata. Unstratified, $IV_a$ and 2SLS are all overplotted, as their performances are essentially the same. Top row shows one-sided noncompliance. Bottom row shows two-sided noncompliance.}
  \label{fig:pred_C_only}
\end{figure}

To explore the tension in strata with few compliers between their increased instability and reduced weight, we conducted a second set of simulations where we generated a series of datasets that all had the same overall compliance rate (15\%), but a series of covariates predictive of compliance that ranged from the compliers being evenly distributed across the strata (no prediction) to all the compliers being in the same stratum (perfect prediction).
All scenarios also had substantial treatment heterogeneity across strata, with $CACE=0$ in the lowest strata and a $CACE$ of $4\sigma$ in the fourth.
We ran two sets of simulations, one with one-sided noncompliance and one with two-sided.
Results are in Figure~\ref{fig:pred_C_only}.

The middle plots of Figure~\ref{fig:pred_C_only} shows the true SEs of the different estimators on the $y$ axis.
The $x$-axis shows the $R^2$ of a regression of compliance onto the strata variable.
Even when we have near 100\% $R^2$ (perfect prediction of compliance type) the $IV_a$ and 2SLS estimators have virtually no gains.
For $IV_w$ and DSS0, we do see precision gains but only at very high values of $R^2$, with DSS0 having slightly larger gains (note $IV_w$ and DSS0 coincide in the one-sided noncompliance setting).
The DSF estimator, more aggressive than $IV_w$ in dropping strata, more quickly realizes gains in the standard errors.
The PWIV estimator has an even sharper drop off in variance, but also is the most biased (see left hand column of plots).

Figure~\ref{fig:pred_C_only} also allows for comparing all estimators to an oracle (the bottom line) of the simple difference in means of the known compliers.
This estimator could be achieved if we knew the complier status of all units.
Such knowledge would be extremely beneficial; the oracle has a standard error less than half of the unstratified estimator.
For one-sided noncompliance, only when we have perfect prediction do $IV_w$, DSF, and PWIV converge to this oracle.
For two-sided noncompliance, if we are not actively discounting or dropping the strata with always-takers, we do not as easily obtain the benefits of our predictor: random imbalances of always-takers makes the other strata still have weight, which introduces instability into the overall estimate.
The DSF and PWIV, by dropping and down-weighting these strata, eventually achieve the oracle's performance.

When we have even only a few compliers across all strata, we immediately are faced with countervailing forces: on one hand, we can down-weight those strata that we know represent a small fraction of the compliers.
On the other hand, the low compliance rates in those strata generate extremely unstable estimates, and so weighting by estimated number of compliers still allow those extreme values to destabilize the overall weighted average.
When we \emph{drop} small strata, however, this destabilization does not occur, and we see benefits.
PWIV's down-weighting unstable strata more heavily here is a type of ``soft dropping.''

Dropping or re-weighting strata does open the door for bias, however.
In our simulation, the treatment impact is higher for compliers in the higher strata, and for scenarios where the compliers are concentrated in the higher strata, the lower strata get dropped either by random 0 estimates of compliance rate (for $IV_w$ and DSS0) or when failing the F-test (for DSF), thus inflating the estimated treatment impact.
For PWIV, the re-weighting of strata introduces bias when there is treatment effect heterogeneity across strata because we are no longer weighting to the sample average, but rather the precision weighted average.
However, the bias is not overly large: the PWIV line in the RMSE plot remains firmly below the other estimators, showing bias is only a fraction of overall error.
For at least these simulations, the drawbacks of introducing bias appear to be more than offset by the benefits of increasing precision.

It is also worth noting in these simulations that even when using a covariate only minimally predictive of compliance type, the RMSE of the post-stratification estimators is no worse than the unstratified estimator (with the exception of DSF which is slightly worse in the two-sided case).
This is further illustrated in the supplementary materials, which show that precision gains can even be achieved when stratifying on covariates unrelated to outcome or compliance type, though there is still a bias tradeoff.
This reassures us that even if in practice we have limited understanding about which covariates are predictive of compliance type, the performance of most of the post-stratification methods will likely be no worse than the unstratified estimator.
This is similar to the finding that as long as a stratification covariate does not lead to more heterogeneity within blocks than between blocks in block randomized experiments, we are unlikely to reduce precision by using a blocked over an unblocked experiment \citep[see,][for more]{pashley2020block}.

\subsection{Discussion}

In our simulations, we considered two properties, prediction of the outcome and prediction of compliance type, of variables that can be used for post-stratification. Post-stratifying can directly improve precision when using covariates predictive of outcome. To gain from a complier-predictive covariate, however, we have to do more than post-stratify: we have to down-weight or drop the ``empty'' or low compliance rate strata. The stability gained from dropping such strata is general: in fact, as we show in the supplementary materials, precision gains can be achieved even if stratifying on covariates completely unrelated to compliance type or outcome, although this comes with associated bias gains.
The standard errors of the post-stratified estimators tend to be relatively well calibrated for one-sided but not two-sided noncompliance, with the delta approach---that nominally takes into account all forms of uncertainty---performing notably worse than Bloom approach for many of the contexts we explored.

These results imply that to exploit a variable predictive of compliance type, analysts need to create multiple strata, and some of those strata need to have a small fraction of compliers. A stratification variable of this type provides a better chance of having estimated compliance rates of zero or close to zero in some strata, which is where the gains come from. As such, a single binary variable is unlikely to be useful unless it can nearly perfectly separate compliers and noncompliers.
One strategy could be to combine multiple binary variables to create many strata, which may result in some strata with few compliers in them.

\section{GOTV Applications}

Next, we employ our post-stratification methods to re-analyze data from two GOTV experiments. In our first analysis, we use data from an RCT designed to compare the effectiveness of three methods of voter contact: door-to-door canvassing, phone calls, and sending mailers \citep{Gerber:2000}. This RCT was conducted in New Haven, Connecticut ahead of the November 1998 election. Households with one or two registered voters were randomized to receive some combination (or none) of the three practices. To simplify the analysis, we focus on door-to-door canvasing, where households either received face-to-face contact from canvassers encouraging them to vote, i.e. treatment, or were not contacted, i.e. control. The outcome of interest is whether either household member voted in the 1998 election. In addition to outcome and treatment assignment, the data also contain covariates on household members such as age and whether they voted in the 1996 election. We use the version of the data \cite{gg_gotv_data} used in \cite{hansen2009attributing}.

In the original analysis, the analysts estimated the CACE using two-stage regression \citep{Gerber:2000}. We follow the original analysis and do not consider the exposure as time varying.
This is reasonable given our focus on the in-person canvasing condition, though might be a concern in the other conditions.
For mailers, the investigators varied how many fliers (up to three) were sent to individuals.
Further, in the phone call condition, multiple contact attempts were made if the initial contact was not successful.
Exploring the use of our methodology with time-varying exposure, which typically makes standard instrumental variable methods perform poorly \citep{Hernan:2006}, would be an interesting direction for future exploration.

We post-stratify based on age, vote in 1996, and household size. Age and household size are likely predictors of being a complier (in this case, being home and answering the door if assigned to door-to-door canvasing), and thus should serve as complier-predictive covariates. Prior vote behavior is likely predictive of future vote behavior (the outcome), and thus is expected to improve precision via either post-stratification or two-stage least squares. In our analysis, we computed the average age by household and split the variable into four approximately equal size groups. Vote in 1996 is defined here as whether either household member voted in two-voter households. This resulted in 23,450 households across 17 strata defined as all realized combinations of these covariates plus a strata for those with missing block variables. Estimated compliance rates across the strata range from about 16\% to 43\%, providing some indication that we are stratifying in a meaningful way in terms of compliance types. Estimated CACEs similarly varied across strata from about -0.19 to 0.43.

Table~\ref{tab:outcomes1} contains CACE estimates based on an unstratified estimator and the proposed stratification methods.
We see an approximately 14\% reduction in standard errors from the post-stratified estimators, except for PWIV, which has approximately an 18\% reduction. Put another way, this means an experiment with only 75\% of the sample size, but using post-stratification, would have similar levels of power to the original; this is a substantial reduction in total effort. This indicates appreciable gains in precision by using post-stratification: we are taking advantage of the complier-predictive nature of our covariates. Similar results in terms of standard error reduction with post-stratification were found using a factorial analysis on the full range of treatment options estimating the Marginal Average Complier Effect, as defined in \cite{blackwellpashley}.
Except for PWIV, the different post-stratification methods result in the same point and variance estimates because no stratum has an estimated 2\% or fewer compliers and no stratum failed the F-test. This illustrates that the extra advantages of the stratification methods that drop strata with low compliance can only be realized when strata are small enough and correlated enough with compliance type to result in low compliance strata. We expect PWIV to introduce potential bias for estimating the average effect among compliers, moving the estimator to target those individuals who are more likely to comply. In this example, this means up-weighting individuals who are older, have previously voted, and come from larger households.
The potential bias will be larger if the effect of door-to-door canvasing among those individuals differs from the effect among the types of individuals who are less likely to comply. For example, it is plausible that individuals who have voted previously are likely to vote again, whether they receive a canvaser or not, leading to smaller estimated effects when those individuals are up-weighted.

\begin{table}[ht]
\centering
\caption{Results from unstratified and various post-stratified IV estimators for door-to-door canvassing GOTV experiment.  Bloom SEs reported.  ``\% SE'' is percent change of SE relative to Unstratified}
\label{tab:outcomes1}
\begin{tabular}{lrrrrrr}
  \toprule
Method & $\widehat{\pi}_c$  & $\widehat{CACE}$ & $\widehat{SE}$ & \% SE & $n$ & p-value\\ 
  \midrule
UNSTRAT & 0.296 & 0.084 & 0.0275 & 100 & 23,450 & 0.0024 \\ 
  $IV_w$ & 0.298 & 0.095 & 0.0238 & 86.2 & 23,450 & 0.0001 \\ 
  $IV_a$ & 0.298 & 0.095 & 0.0238 & 86.2 & 23,450 & 0.0001 \\ 
  DSS (2\%) & 0.298 & 0.095 & 0.0238 & 86.2 & 23,450 & 0.0001 \\ 
  PWIV  & 0.298 & 0.092 & 0.0226& 82.0& 23,450 & 0.0000 \\ 
  DSF  & 0.298 & 0.095 & 0.0238 & 86.2 & 23,450 & 0.0001 \\
   \bottomrule
\end{tabular}
\end{table}

The second study we analyze is the evaluation of door-to-door canvasing described in the introduction \citep{Green:2003a}. Replication data is provided in \citet{gotv_data1}. In the study, 8,580 votes were assigned to the treatment condition and 10,081 we assigned to control. Turfs (randomization blocks) ranged in size from 16 voters to 535 voters, with an average size of 124 voters. We use these 150 turfs as strata, since compliance rates differed significantly across each turf. We find the estimated compliance rates varied from 5\% to 69\% with an average compliance rate of 30\%. Moreover, turnout rates also varied substantially by turf (estimates from a multilevel logistic model suggest true voting rates ranged from 6\% to 61\%), showing this stratification covariate is also predictive of the outcome. 

Table~\ref{tab:outcomes2} contains CACE estimates based on an unstratified estimator and the proposed stratification methods. The unstratified estimate indicates that exposure to canvassing increased voter turnout by 8\%. The estimates from the stratified methods are all between 4 and 5\%, nearly 50\% smaller. In this case the precision gains are more modest, with reductions of around 4\%. There were no strata with estimated 2\% proportion of compliers or less, leading to $IV_w$, $IV_a$, and DSS having the same estimates. However, not all strata passed the F-test, so DSF results in a different estimate. One interesting feature of the analysis is that, while the DSF estimator had a notable reduction in sample size (close to 15\% reduction), the standard error estimate is not any larger than for the unstratified estimator. This emphasizes that for IV estimates, not all data are useful.

\begin{table}[ht]
\centering
\caption{Results from unstratified and various post-stratified IV estimators for door-to-door canvassing GOTV experiment.  Bloom SEs reported.  ``\% SE'' is percent change of SE relative to Unstratified}
\label{tab:outcomes2}
\begin{tabular}{lrrrrrr}
  \toprule
Method & $\widehat{\pi}_c$  & $\widehat{CACE}$ & $\widehat{SE}$ & \% SE & $n$ & p-value\\ 
  \midrule
UNSTRAT & 0.295 & 0.082 & 0.0228 & 100 & 18,661 & 0.00 \\ 
  $IV_w$ & 0.300 & 0.055 & 0.0219 & 96.0 & 18,661 & 0.01 \\ 
  $IV_a$ & 0.300 & 0.055 & 0.0219 & 96.0 & 18,661 & 0.01 \\ 
  DSS (2\%) & 0.300 & 0.055 & 0.0219 & 96.0 & 18,661 & 0.01 \\ 
  PWIV & 0.300 & 0.042 & 0.0197 & 86.1 & 18,661 & 0.03 \\ 
  DSF & 0.323 & 0.047 & 0.0222 & 97.1 & 16,010 & 0.03 \\
   \bottomrule
\end{tabular}
\end{table}

\section{Conclusion}
 
We have explored the benefits of combining IV estimators with post-stratification. We outlined the gains that are possible both analytically and through a series of simulations. We largely focused on how to use post-stratification to take advantage of a baseline covariate predictive of compliance behavior, rather than the outcome. Classic IV estimation methods are not designed to exploit the information from covariates of this type. Post-stratification on these covariates, by isolating compliers in some strata and dropping the rest, can, however, provide important precision gains. We also studied the post-stratified IV estimator more broadly. The theoretical advantages of this approach include lower bias, lower variance, and lower variability of the SE estimates, especially when stratifying on a covariate predictive of outcome.

In practice, researchers will need to make choices on how to stratify and which method to use for estimation with post-stratification.
While stratifying on covariates predictive of outcome directly boosts precision, stratifying on covariates predictive of compliance type only improves precision if we are able to drop or down-weight lower compliance strata. Therefore, when considering which covariates to use in post-stratification with IV, it is advantageous that covariates believed to be predictive of compliance type are not too coarse (e.g., not binary), or that several such covariates are combined.
Such a finer grain stratification allows, in principle, greater variation of complier proportions across the strata, increasing the likelihood that we can drop strata with no or very few compliers.
That said, there is a trade-off with strata that are too fine: We risk ending up with one or fewer treated or control units in a stratum, making inference for that stratum infeasible. It would therefore be advantageous for researchers to pick stratification variables that yield large enough groups that the risk of a singleton treated or control unit within a stratum is very low. This may mean focusing on covariates predictive of outcome when covariates predictive of compliance type would be too coarse.

To maintain the validity of inferences after post-stratification, researchers should choose stratification variables and the method of stratification prior to seeing the data by relying on prior studies or subject matter knowledge to choose covariates, or possibly through sample splitting, ideally as detailed in a publicly available pre-analysis plan.
Researchers should also state their intended method for evaluation prior to seeing the data, including the criteria for droppings strata if using methods such as DSS or DSF.
Overall, our results show that $IV_w$ (and in the two-sided case, DSS0 which also drops strata with negative estimated compliers) should generally be preferred to $IV_a$, and that methods that further drop strata or down-weight low-compliance strata, such as PWIV, may be even more preferable based on a researcher's tolerance of the bias-variance tradeoff.

One limitation of the methods presented here is that they focus on complier effects and, through dropping or down-weighting strata, may become even more local.
Therefore, generalizations of results to other populations may not be plausible and at a minimum should be made with a degree of skepticism. Ideally, IV analyses as described here contribute evidence of effects that either lead to further experiments that are able to better control compliance or can be combined in a meta-analysis to provide a more complete picture of the treatment effect. \citet[p. 522]{CausalInferenceText} note that when reasons for noncompliance are idiosyncratic to the setting in which the experiment is conducted, local effects may have greater external validity than intention to treat effects.  As such, the more local effects outlined in our work may be compelling in some applications.

As noted above, IV post-stratification methods have the widest applicability in contexts where the instrument passes the classic weak instrument test but compliance is relatively low.
Such a pattern is typical for RCTs on the effectiveness of GOTV methods, where difficulties of directly contacting voters by any method generally induces relatively low levels of compliance.
Other areas of research that use RCTS face similar compliance patterns.
For example, the evaluation of youth education and outreach efforts also often have low levels of compliance \citep{guryan2023not, heller2014summer}.

To use these approaches, analysts needs to identify a variable that allows stratifying units into strata with more or fewer compliers (and/or systematically different levels of outcome).
Ideally, post-stratification can be built into the design phase of the experiment.
Collection of such data could be a part of the design phase of future GOTV RCTs.
In addition, better stratification variables might be identified in the future. GOTV RCTs are regularly conducted during election cycles \citep{green2019get}. Future iterations of GOTV RCTs could be used to identify additional variables for post-stratification.
More generally, data from any similar experimental design could be used to identify post-stratification variables for use in future experiments.

\clearpage

\bibliographystyle{apalike}
\bibliography{iv_ref,gotv}{}

\begin{appendices}
\include{iv_post_append}

\end{appendices}

\end{document}

%% file: iv_post_append.tex
\newpage
\setcounter{page}{1}
\begin{center}
{\bf \Large  Supplementary Material\\for\\``Improving instrumental variable estimators with post-stratification''}
\end{center}

\section{Code and replication files}
An R package to implement all the post-stratified instrumental variable methods discussed in the paper can be found at \url{https://github.com/lmiratrix/poststratIV}.
This repository also includes replication files for all simulations and the GOTV applications.

\section{Equivalence with weighted Two-Stage Least Squares}\label{append:2sls}
When performing a generic weighted least squares regression of $y$ and $x$ ($y_i = \beta_0 + \beta_1x_i + \epsilon_i$) with weights $w$, we have the following formulas for the regression estimates:
\begin{align}
\hat{\beta}_1 &= \frac{\sum_{i=1}^nw_i(y_i-\overline{y}_w)(x_i - \overline{x}_w)}{\sum_{i=1}^nw_i(x_i - \overline{x}_w)^2}\label{eq:beta_1}\\ 
\hat{\beta}_0 &= \overline{y}_w  - \hat{\beta}_1\overline{x}_w \nonumber
\end{align}
where $\overline{y}_w = \sum_{i=1}^nw_iy_i/\sum_{i=1}^nw_i$ and $\overline{x}_w$ is defined analogously.

\subsection{First stage results}
\begin{result}
Consider performing a weighted regression of $D$ on $Z$ ($D_i = \beta_{0,S1} + \beta_{1,S1}Z_i + \epsilon_i$) with weights $w_i = \frac{N_{g}}{N_{g,z}}\frac{n_z}{N}$ for unit $i$ in strata $g \in \{1,\dots,G\}$ assigned to treatment $z \in \{0,1\}$.
Then our estimate for coefficient for $Z$ is $\hat{\beta}_{1, S1} =\widehat{f}_{PS}$ and for the intercept it is $\hat{\beta}_{0, S1} =\overline{D}_{w,0}^{obs}$.
The predicted values from this model are $D^{\text{pred}}_i = Z_i\overline{D}_{w,1}^{obs} + (1-Z_i)\overline{D}_{w,0}^{obs}$.
\end{result}

\begin{proof}
That the coefficient from the model described will be equivalent to the blocked treatment effect estimator is an established result \citep[see, e.g., ][]{pashley2021insights} but we will here provide a proof and derive the predicted values for completeness.

First, we will find the weighted means for $D$ and $Z$.
\begin{align*}
\overline{Z}_w &= \sum_{i=1}^nw_iZ_i/\sum_{i=1}^nw_i \\
&= \left[\sum_{g=1}^G\sum_{i:s_i = g}Z_i\frac{N_{g}}{N_{g,1}}\frac{n_1}{N}\right]/\left[\sum_{g=1}^G\sum_{i:s_i = g}\left(Z_i\frac{N_{g}}{N_{g,1}}\frac{n_1}{N} + (1-Z_i)\frac{N_{g}}{N_{g,0}}\frac{n_0}{N}\right)\right]\\
&= \left[\sum_{g=1}^GN_{g}\frac{n_1}{N}\right]/\left[\sum_{g=1}^G\left(N_{g}\frac{n_1}{N} + N_{g}\frac{n_0}{N}\right)\right]\\
&= n_1/N
\end{align*}

\begin{align*}
\overline{D}_w &= \sum_{i=1}^nw_iD_i/\sum_{i=1}^nw_i \\
&=\frac{1}{N} \sum_{g=1}^G\sum_{i:s_i = g}\left[Z_iD_i(1)\frac{N_{g}}{N_{g,1}}\frac{n_1}{N} + (1-Z_i)D_i(0)\frac{N_{g}}{N_{g,0}}\frac{n_0}{N}\right]\\
&=\frac{1}{N} \sum_{g=1}^G\left[\overline{D}_{g,1}^{obs}(N_{g}\frac{n_1}{N} + \overline{D}_{g,0}^{obs}N_{g}\frac{n_0}{N}\right]\\
&=\frac{n_1}{N}\overline{D}_{w,1}^{obs} + \frac{n_0}{N}\overline{D}_{w,0}^{obs},
\end{align*}
where $\overline{D}_{w,z}^{obs} = \sum_{g=1}^G\frac{N_g}{N}\overline{D}_{g,z}^{obs}$ for $z \in \{0,1\}$.

Then for the denominator of $\hat{\beta}_{1, S1}$, following Equation~(\ref{eq:beta_1}), we have
\begin{align*}
\sum_{i=1}^nw_i(Z_i - \overline{Z}_w)^2 &=  \sum_{g=1}^G\sum_{i:s_i = g}\Big[Z_i\frac{N_{g}}{N_{g,1}}\frac{n_1}{N} \left(1 - n_1/N\right)^2 + (1-Z_i)\frac{N_{g}}{N_{g,0}}\frac{n_0}{N} \left( -  n_1/N\right)^2\Big]\\
&=  \sum_{g=1}^G\Big[N_{g}\frac{n_1}{N} \frac{n_0^2}{N^2} + N_{g}\frac{n_0}{N}\frac{n_1^2}{N^2}\Big]\\
&=  \frac{n_1n_0}{N}.
\end{align*}

For the numerator we have 
\begin{align*}
&\sum_{i=1}^nw_i(D_i - \overline{D}_w)(Z_i - \overline{Z}_w)\\
 &=  \sum_{g=1}^G\sum_{i:s_i = g}\Big[Z_i\frac{N_{g}}{N_{g,1}}\frac{n_1}{N} \left(D_i(1) - \frac{n_1}{N}\overline{D}_{w,1}^{obs} - \frac{n_0}{N}\overline{D}_{w,0}^{obs}\right)\left(1 - n_1/N\right) \\
&\qquad \qquad  \qquad+ (1-Z_i)\frac{N_{g}}{N_{g,0}}\frac{n_0}{N} \left(D_i(0) - \frac{n_1}{N}\overline{D}_{w,1}^{obs} - \frac{n_0}{N}\overline{D}_{w,0}^{obs}\right)\left( -  n_1/N\right)\Big]\\
&= \frac{n_1n_0}{N^2} \sum_{g=1}^GN_g\sum_{i:s_i = g}\Big[\frac{Z_i}{N_{g,1}} \left(D_i(1) - \frac{n_1}{N}\overline{D}_{w,1}^{obs} - \frac{n_0}{N}\overline{D}_{w,0}^{obs}\right)\\
&\qquad \qquad  \qquad  \qquad  \qquad - \frac{1-Z_i}{N_{g,0}} \left(D_i(0) - \frac{n_1}{N}\overline{D}_{w,1}^{obs} - \frac{n_0}{N}\overline{D}_{w,0}^{obs}\right)\Big]\\
&= \frac{n_1n_0}{N^2} \sum_{g=1}^GN_g\Big[ \overline{D}_{g,1}^{obs} - \frac{n_1}{N}\overline{D}_{w,1}^{obs} - \frac{n_0}{N}\overline{D}_{w,0}^{obs}- \overline{D}_{g,0}^{obs} + \frac{n_1}{N}\overline{D}_{w,1}^{obs} + \frac{n_0}{N}\overline{D}_{w,0}^{obs}\Big]\\
&= \frac{n_1n_0}{N^2} \sum_{g=1}^GN_g\Big[ \overline{D}_{g,1}^{obs} - \overline{D}_{g,0}^{obs} \Big]\\
&= \frac{n_1n_0}{N}\widehat{f}_{PS}.
\end{align*}

All together,
\[\hat{\beta}_{1, S1} =\widehat{f}_{PS}. \]

The intercept is 
\[\hat{\beta}_0 =\frac{n_1}{N}\overline{D}_{w,1}^{obs} + \frac{n_0}{N}\overline{D}_{w,0}^{obs} - \frac{n_1}{N}\widehat{f}_{PS} =\overline{D}_{w,0}^{obs} .\]

The predicted values are 
\[{D}^{\text{pred}}_i = Z_i\widehat{f}_{PS} +\overline{D}_{w,0}^{obs} = Z_i\overline{D}_{w,1}^{obs} + (1-Z_i)\overline{D}_{w,0}^{obs}. \]

\end{proof}

\subsection{Second stage results}

\begin{result}
Lemma~\ref{lemma:2sls_connection}:
Consider performing the second-stage weighted regression of $Y$ on $D^{pred}$  ($Y_i = \beta_{0,S2} + \beta_{1,S2}D^{\text{pred}}_i + \epsilon_i$) with weights $w_i = \frac{N_{g}}{N_{g,z}}\frac{n_z}{N}$ for unit $i$ in strata $g \in \{1,\dots,G\}$ assigned to treatment $z \in \{0,1\}$.
Then our estimate of the coefficient for $D^{pred}$ is $\hat{\beta}_{1, S2} = \widehat{\CACE}_{\text{IV-a}}$.
\end{result}

\begin{proof}
Following the same logic we used to find $\overline{D}_w$, the weighted mean for the outcomes is
\begin{align*}
\overline{Y}_w 
&=\frac{n_1}{N}\overline{Y}_{w,1}^{obs} + \frac{n_0}{N}\overline{Y}_{w,0}^{obs},
\end{align*}
where $\overline{Y}_{w,z}^{obs} = \sum_{g=1}^G\frac{N_g}{N}\overline{Y}_{g,z}^{obs}$ for $z \in \{0,1\}$.

For the predicted uptake, the weighted mean is
\begin{align*}
\overline{D}_w^{pred} 
&=\frac{n_1}{N}\overline{D}_{w,1}^{obs} + \frac{n_0}{N}\overline{D}_{w,0}^{obs} = \overline{D}_w.
\end{align*}

Using Equation~(\ref{eq:beta_1}), the denominator of $\hat{\beta}_{1, S2}$ is
\begin{align*}
&\sum_{i=1}^nw_i(D_i^{pred} - \overline{D}_w)^2 \\
&=  \sum_{g=1}^G\sum_{i:s_i = g}\Big[Z_i\frac{N_{g}}{N_{g,1}}\frac{n_1}{N} \left(\overline{D}_{w,1}^{obs} - \frac{n_1}{N}\overline{D}_{w,1}^{obs} - \frac{n_0}{N}\overline{D}_{w,0}^{obs}\right)^2\\
& \qquad \qquad \qquad  + (1-Z_i)\frac{N_{g}}{N_{g,0}}\frac{n_0}{N} \left(\overline{D}_{w,0}^{obs} - \frac{n_1}{N}\overline{D}_{w,1}^{obs} - \frac{n_0}{N}\overline{D}_{w,0}^{obs}\right)^2\Big]\\
&=  \sum_{g=1}^G\sum_{i:s_i = g}\Big[Z_i\frac{N_{g}}{N_{g,1}}\frac{n_1}{N}\frac{n_0^2}{N^2} \left(\overline{D}_{w,1}^{obs} - \overline{D}_{w,0}^{obs}\right)^2  + (1-Z_i)\frac{N_{g}}{N_{g,0}}\frac{n_0}{N}\frac{n_1^2}{N^2} \left(\overline{D}_{w,1}^{obs} - \overline{D}_{w,0}^{obs}\right)^2\Big]\\
&=  \frac{n_1n_0}{N} \left(\overline{D}_{w,1}^{obs} - \overline{D}_{w,0}^{obs}\right)^2\\
& = \frac{n_1n_0}{N}\widehat{f}_{PS}^2.
\end{align*}

For the numerator we have
\begin{align*}
&\sum_{i=1}^nw_i(D_i^{pred} - \overline{D}_w)(Y_i - \overline{Y}_w) \\
&=  \sum_{g=1}^G\sum_{i:s_i = g}\Big[Z_i\frac{N_{g}}{N_{g,1}}\frac{n_1}{N}\frac{n_0}{N} \left(\overline{D}_{w,1}^{obs} - \overline{D}_{w,0}^{obs}\right)\left(Y_i(1) - \frac{n_1}{N}\overline{Y}_{w,1}^{obs} - \frac{n_0}{N}\overline{Y}_{w,0}^{obs}\right) \\
& \qquad \qquad \qquad  - (1-Z_i)\frac{N_{g}}{N_{g,0}}\frac{n_0}{N}\frac{n_1}{N} \left(\overline{D}_{w,1}^{obs} - \overline{D}_{w,0}^{obs}\right)\left(Y_i(0) - \frac{n_1}{N}\overline{Y}_{w,1}^{obs} - \frac{n_0}{N}\overline{Y}_{w,0}^{obs}\right)\Big]\\
&= \frac{n_1n_0}{N^2}\widehat{f}_{PS} \sum_{g=1}^GN_g\sum_{i:s_i = g}\Big[\frac{Z_i}{N_{g,1}}\left(Y_i(1) - \frac{n_1}{N}\overline{Y}_{w,1}^{obs} - \frac{n_0}{N}\overline{Y}_{w,0}^{obs}\right)\\
& \qquad \qquad \qquad  \qquad \qquad \qquad \qquad  - \frac{1-Z_i}{N_{g,0}}\left(Y_i(0) - \frac{n_1}{N}\overline{Y}_{w,1}^{obs} - \frac{n_0}{N}\overline{Y}_{w,0}^{obs}\right)\Big]\\
&= \frac{n_1n_0}{N^2}\widehat{f}_{PS} \sum_{g=1}^GN_g\Big[\left(\overline{Y}_{g,1}^{obs} - \frac{n_1}{N}\overline{Y}_{w,1}^{obs} - \frac{n_0}{N}\overline{Y}_{w,0}^{obs}\right) - \left(\overline{Y}_{g,0}^{obs} - \frac{n_1}{N}\overline{Y}_{w,1}^{obs} - \frac{n_0}{N}\overline{Y}_{w,0}^{obs}\right)\Big]\\
&= \frac{n_1n_0}{N^2}\widehat{f}_{PS} \sum_{g=1}^GN_g\Big[\overline{Y}_{g,1}^{obs} - \overline{Y}_{g,0}^{obs}\Big]\\
&= \frac{n_1n_0}{N}\widehat{f}_{PS}\widehat{\ITT}_{\text{PS}}.
\end{align*}

Putting it together, we have $\hat{\beta}_{1, S1} = \widehat{\ITT}_{\text{PS}}/\widehat{f}_{PS} = \widehat{\CACE}_{\text{IV-a}}$.

\end{proof}

\section{IV variance via delta method}\label{append:iv_var}
In this section, we show how the treatment uptake variance and CLT conditions simplify to those giving in the paper.

\subsection{Simplification of variance and CLT conditions for treatment uptake}\label{app:var_cond_simple}
We first show the simplified variance expressions we use in Section~\ref{sec:standard_iv_est}.
We have the following simplifications of variances for treatment uptake:
\begin{align*}
S^2_D(1) &= \frac{1}{N-1}\sum_{i=1}^N(D_i(1) - \overline{D}(1))^2\\
&= \frac{1}{N-1}\left(N\pi_{n}(\pi_{c} + \pi_{a})^2 + N(\pi_{c} + \pi_{a})\pi_{n}^2\right)\\
&= \frac{N}{N-1}\pi_{n}(\pi_{c} + \pi_{a})
\end{align*}
\begin{align*}
S^2_D(0) &= \frac{1}{N-1}\sum_{i=1}^N(D_i(0) - \overline{D}(0))^2\\
&= \frac{1}{N-1}\left(N\pi_{a}(\pi_{c} + \pi_{n})^2 + N(\pi_{c} + \pi_{n})\pi_{a}^2\right)\\
&= \frac{N}{N-1}\pi_{a}(\pi_{c} + \pi_{n})
\end{align*}
and
\begin{align*}
S^2_D(01) &= \frac{1}{N-1}\sum_{i=1}^N(D_i(1) -D_i(0)  - \overline{D}(1) + \overline{D}(0))^2\\
&= \frac{1}{N-1}\left(N\pi_{c}(\pi_{a} + \pi_{n})^2 + N(\pi_{a} + \pi_{n})\pi_{c}^2 \right)\\
&= \frac{N}{N-1}\pi_{c}(\pi_{a} + \pi_{n}).
\end{align*}

We also have 
\[v_{D}(z) = \max_{1\leq i \leq N}\left(D_i(z) - \overline{D}(z)\right)^2\]
where
\begin{align*}
\max_{1\leq i \leq N}\left(D_i(1) - \overline{D}(1)\right)^2 = \max \left(\mathbb{I}\left(\pi_n \neq 0\right)(\pi_{c} + \pi_{a})^2, \mathbb{I}\left(\pi_n \neq 1\right)\pi_{n}^2\right)
\end{align*}
and
\begin{align*}
\max_{1\leq i \leq N}\left(D_i(0) - \overline{D}(0)\right)^2 = \max \left(\mathbb{I}\left(\pi_a \neq 0\right)(\pi_{c} + \pi_{n})^2, \mathbb{I}\left(\pi_a \neq 1\right)\pi_{a}^2\right)
\end{align*}

The condition from \cite{LiDin17} (Theorem 4) to obtain a finite-population central limit result for $\hat{f}$ is
\begin{align*}
Q_n \coloneqq \max_{z \in\{0,1\}}\frac{1}{n_z^2}\frac{v_{D}(z)}{n_0^{-1}S^2_D(0) + n_1^{-1}S^2_D(1) - N^{-1}S^2_D(01)} \to 0
\end{align*}
but
\begin{align*}
Q_n < \frac{1}{\min(p, 1-p)N}\frac{1}{ \frac{N}{N-1}\pi_{a}(\pi_{c} + \pi_{n}) + p^{-1}\frac{N}{N-1}\pi_{n}(\pi_{c} + \pi_{a}) - \frac{N}{N-1}\pi_{c}(\pi_{a} + \pi_{n})}.
\end{align*}

This is clearly satisfied if $\pi_{c}$, $\pi_{a}$, and $\pi_{n}$ have asymptotic limiting values such that at least two of those proportions are bounded away from zero.
As the second term will go to 0 as $N \to \infty$, we have our condtion.
Therefore, if two of $\pi_{c}$, $\pi_{a}$, and $\pi_{n}$ are nonzero, $\hat{f}$ is asymptotically normal.

\subsection{Finite-population CLT assumptions}\label{append:clt_post_iv}

In order to consider asymptotic variance, we need to fix an asymptotic regime to work within.
Here, we will assume a finite number of strata, with the size of each growing to infinity, as outlined in the following assumption.
\begin{assumption}\label{assump:strata_prop}
The number of strata $G$ is fixed and the number of units $N_g$ grows as $N \to \infty$, such that $N_g/N \to h_g$, where $h_g \in (0,1)$ is some constant.
\end{assumption}

We will use various additional assumptions for the CLTs of the post-stratified estimators:
\begin{assumption}\label{assump:li_ding_cond_main}
Define $c_g(z) =\max_{1 \leq i \leq N} \frac{N^2}{N_g^2}W_i(g)\left(Y_i(z) - \overline{Y}_g(z)\right)^2$,
\[v(z) = \frac{1}{N-1}\sum_{g=1}^G\sum_{i:s_i=g}\left(Y_i(z) - \overline{Y}_g(z)\right)^2,\]
and
\[v(01)= \frac{1}{N-1}\sum_{g=1}^G\sum_{i:s_i=g}\left(Y_i(1) - Y_i(0) -\left[ \overline{Y}_g(1) - \overline{Y}_g(0) \right]\right)^2\]
As $N \to \infty$
\[\max_{z \in {0,1}}\max_{g \in \{1,\dots,G\}}\frac{1}{n_z^2}\frac{c_g(z) }{n_0^{-1}v(0) + n_1^{-1}v(1) - N^{-1}v(01)} \to 0.\]
\end{assumption}
\begin{assumption}\label{assump:clt_cond2_main}
$S^2_{g,Y}(z)$ and $S^2_{g,Y}(1,0)$ have finite limiting values as $N \to\infty$.
\end{assumption}
\begin{assumption}\label{assump:delta_post_strat}
$N\text{var}(\widehat{\ITT}_{\text{PS}})$ has a finite limiting value, to help ensure $\widehat{\ITT}_{\text{PS}} - \ITT \overset{p}{\to} 0 $.
\end{assumption}

\subsection{CLT results for post-stratification ITT estimators}\label{append:clt_itt}
We first show that both $\widehat{\ITT}_{\text{PS}}$ and $\widehat{f}_{\text{PS}}$ have asymptotic normal distributions.

\begin{lemma}\label{lem:post_strat_clt_y}
If we have Assumptions \ref{assump:iv} (Part A), \ref{assump:strata_prop}, \ref{assump:li_ding_cond_main}, and \ref{assump:clt_cond2_main} then by results from \cite{schochet2023design},
\begin{align*}
 \frac{\widehat{\ITT}_{\text{PS}} - \ITT}{\sqrt{\text{asyVar}(\widehat{\ITT}_{\text{PS}})}} \overset{d}{\to} N(0,1).
\end{align*}
where
\begin{align*}
\text{asyVar}\left(\widehat{\ITT}_{\text{PS}}\right) 
& =  \sum_{g=1}^N\frac{N_g}{N}\frac{N_g-1}{N-1}\left[\frac{S^2_{g,Y}(0)}{(1-p)N_g} +\frac{S^2_{g,Y}(1)}{pN_g} - \frac{S^2_{g, Y}(01)}{N_g}\right]\\
& \approx  \sum_{g=1}^N\frac{N_g^2}{N^2}\left[\frac{S^2_{g,Y}(0)}{(1-p)N_g} +\frac{S^2_{g,Y}(1)}{pN_g} - \frac{S^2_{g, Y}(01)}{N_g}\right].
\end{align*}
\end{lemma}
We note that we simplify some of the conditions from \cite{schochet2023design} because (i) we do not have additional covariate adjustment (ii) the weights in our setting correspond to the post-stratification weighting of units, and Assumption~\ref{assump:strata_prop} sets limits on the asymptotic behavior of those weights.

For $\widehat{f}_{\text{PS}}$ we need an extension of above:
\begin{assumption}\label{assump:clt_cond_d}
Assume that $\pi_{g,c}$, $\pi_{g,a}$, and $\pi_{g,n}$ have limiting values as $N \to \infty$.
Also assume at least two of $\pi_{g,c}$, $\pi_{g,a}$, and $\pi_{g,n}$ are asymptotically bounded away from zero for at least one $g \in \{1,\dots,G\}$.
\end{assumption}

\begin{lemma}\label{lem:post_strat_clt_d}
If we have Assumptions~\ref{assump:iv} (Part A), \ref{assump:strata_prop} and \ref{assump:clt_cond_d} 
then by results from \cite{schochet2023design},
\begin{align*}
\frac{\widehat{f}_{\text{PS}} - \pi_c}{\sqrt{\text{asyVar}(\widehat{f}_{\text{PS}})}} \overset{d}{\to} N(0,1),
\end{align*}
where
\begin{align*}
\text{asyVar}(\widehat{f}_{\text{PS}}) &= \frac{1}{N-1}\sum_{g=1}^GN_g\left[\frac{ \left(\pi_{g,c} + \pi_{g,a}\right)\pi_{g,n} }{n_0} + \frac{\left(\pi_{g,c} + \pi_{g,n}\right)\pi_{g,a}}{n_1} - \frac{1}{N}\left(\pi_{g,a} + \pi_{g,n}\right)\pi_{g,c}\right]\\
&= \sum_{g=1}^G\frac{N_g}{N(N-1)}\left[(1-p)^{-1}\left(\pi_{g,c} + \pi_{g,a}\right)\pi_{g,n} + p^{-1}\left(\pi_{g,c} + \pi_{g,n}\right)\pi_{g,a} - \left(\pi_{g,a} + \pi_{g,n}\right)\pi_{g,c}\right].
\end{align*}
\end{lemma}
In this case, the bounded nature of uptake $D$, along with Assumption~\ref{assump:clt_cond_d}  implies some of the usual conditions for the central limit theorem hold.

\section{Bias reduction}
\subsection{One-sided noncompliance}\label{append:one_sided_bias}
First let $\overline{Y}_c(z)$ be the average potential outcome under treatment $z$ among compliers.
Similarly, let $\overline{Y}_n(z)$ be the average potential outcome under treatment $z$ among noncompliers (never-takers) and note that $\overline{Y}_n(1)=\overline{Y}_n(0) = \overline{Y}_n$.
Let $\widehat{\overline{Y}}_c(z)$ and $\widehat{\overline{Y}}_n(z)$ be the corresponding estimates we would get based on treatment assignment if we could observe who is a complier or noncomplier.
Then we can write
\begin{align*}
\widehat{\ITT} &= \overline{Y}^{\text{obs}}(1) -  \overline{Y}^{\text{obs}}(0)\\
&= \hat{f}\widehat{\overline{Y}}_c(1) + (1-\hat{f})\widehat{\overline{Y}}_n(1) -\frac{n_c - n_1\hat{f}}{n_0}\widehat{\overline{Y}}_c(0) -\frac{n_0 - n_c + n_1\hat{f}}{n_0}\widehat{\overline{Y}}_n(0)\\
&=\hat{f}\left(\widehat{\overline{Y}}_c(1) - \widehat{\overline{Y}}_n +\frac{n_1}{n_0}\widehat{\overline{Y}}_c(0) - \frac{n_1}{n_0}\widehat{\overline{Y}}_n\right)
+\left(\widehat{\overline{Y}}_n - \frac{n_c}{n_0}\widehat{\overline{Y}}_c(0) - \frac{n_0 - n_c}{n_0}\widehat{\overline{Y}}_n\right)\\
&=\hat{f}\left(\widehat{\overline{Y}}_c(1)  +\frac{n_1}{n_0}\widehat{\overline{Y}}_c(0) - \frac{N}{n_0}\widehat{\overline{Y}}_n\right)
+\left( - \frac{n_c}{n_0}\widehat{\overline{Y}}_c(0) + \frac{n_c}{n_0}\widehat{\overline{Y}}_n\right)
\end{align*}

Therefore,
\begin{align*}
E\left[\widehat{\CACE}\right] &= E\left[E\left[\frac{\widehat{\ITT}}{\hat{f}}|\hat{f}\right]\right]\\
&= E\left[\left(\widehat{\overline{Y}}_c(1)  +\frac{n_1}{n_0}\widehat{\overline{Y}}_c(0) - \frac{N}{n_0}\widehat{\overline{Y}}_n\right) + \frac{1}{\hat{f}}E\left[ - \frac{n_c}{n_0}\widehat{\overline{Y}}_c(0) + \frac{n_c}{n_0}\widehat{\overline{Y}}_n|\hat{f}\right]\right]\\
&= \left(\overline{Y}_c(1)  +\frac{n_1}{n_0}\overline{Y}_c(0)  - \frac{N}{n_0}\overline{Y}_n\right) +E\left[ \frac{1}{\hat{f}}\right]\left( - \frac{n_c}{n_0}\overline{Y}_c(0) + \frac{ n_c}{n_0}\overline{Y}_n\right)\\
&= \overline{Y}_c(1) -\left( \frac{N}{n_0} - E\left[ \frac{1}{\hat{f}}\right]\frac{E[\hat{f}]N}{n_0}\right)\overline{Y}_n - \left(E\left[ \frac{1}{\hat{f}}\right]\frac{E[\hat{f}]N}{n_0} - \frac{n_1}{n_0}\right)\overline{Y}_c(0)\\
&= \overline{Y}_c(1) -\frac{1}{1-p}\left(1  - E\left[ \frac{1}{\hat{f}}\right]E[\hat{f}]\right)\overline{Y}_n - \frac{1}{1-p}\left(E\left[ \frac{1}{\hat{f}}\right]E[\hat{f}] - p\right)\overline{Y}_c(0)\\
\end{align*}

The bias of $\widehat{\CACE}_{\text{IV}}$ is then
\begin{align*}
E\left[\widehat{\CACE}_{\text{IV}}\right] - \CACE &=  -\frac{1}{1-p}\left(1  - E\left[ \frac{1}{\hat{f}}\right]E[\hat{f}]\right)\overline{Y}_n - \frac{1}{1-p}\left(E\left[ \frac{1}{\hat{f}}\right]E[\hat{f}] - 1\right)\overline{Y}_c(0)\\
&=  \frac{1}{1-p}\left(1  - E\left[\frac{1}{ \hat{f}}\right]E[\hat{f}]\right)\left( \overline{Y}_c(0)-\overline{Y}_n\right)
\end{align*}
Note that $\text{cov}\left(\hat{f}, \frac{1}{\hat{f}}\right) = 1  - E\left[ \frac{1}{\hat{f}}\right]E[\hat{f}]$, so this piece is always negative.

We can get an estimation of the bias using a Taylor expansion.
To fo this we need to get the moments of $\hat{f}$. Under a completely randomized design with one-sided noncompliance, $\hat{f}$ will follow a hypergeometric distribution.
Using a binomial will give us a reasonable (and simpler) approximation.

\[g(\hat{f}) \approx g(\pi_c) + g'(\pi_c)(\hat{f} -\pi_c) +  \frac{g^{''}(\pi_c)}{2}(\hat{f} - \pi_c)^2 + \frac{g^{'''}(\pi_c)}{3!}(\hat{f} - \pi_c)^3 + \frac{g^{(4)}(\pi_c)}{4!}(\hat{f} - \pi_c)^4. \]
Here we have $g(\hat{f}) = \frac{1}{\hat{f}}$ so
\[\frac{1}{\hat{f}} \approx \frac{1}{\pi_c} - \frac{1}{\pi_c^2}(\hat{f} -\pi_c) +  \frac{1}{\pi_c^3}(\hat{f} - \pi_c)^2 - \frac{1}{\pi_c^4}(\hat{f} - \pi_c)^3 + \frac{1}{\pi_c^5}(\hat{f} - \pi_c)^4. \]
Taking expectations of both sides,
\begin{align*}
E\left[\frac{1}{\hat{f}}\right] &\approx \frac{1}{\pi_c} - 0 +  \frac{1}{\pi_c^3}E[(\hat{f} - \pi_c)^2] - \frac{1}{\pi_c^4}E[(\hat{f} - \pi_c)^3 ]+ \frac{1}{\pi_c^5}E[(\hat{f} - \pi_c)^4]\\
&\text{Using binomial as an approximation for the moments...}\\
&\approx \frac{1}{\pi_c} +  \frac{\pi_c(1-\pi_c)}{Np\pi_c^3} - \frac{\pi_c(1-\pi_c)(1-2\pi_c)}{(Np)^2\pi_c^4}+ \frac{\pi_c(1-\pi_c)(1+(3Np - 6)\pi_c(1-\pi_c))}{(Np)^3\pi_c^5}\\
&\approx \frac{1}{\pi_c} +  \frac{\text{var}(\hat{f})}{\pi_c^3}\left[1 - \frac{1-2\pi_c}{(Np)\pi_c}+ \frac{1+(3Np - 6)\pi_c(1-\pi_c)}{(Np)^2\pi_c^2}\right].
\end{align*}

We see that higher variability of $\hat{f}$ will increase the bias.
Another sensible thing this approximation reveals is that bias will be larger the lower the compliance rate is.

Using hypergeometric moments (based on the completely randomized assignment distribution) instead gives:
\begin{align*}
E\left[\frac{1}{\hat{f}}\right] &\approx \frac{1}{\pi_c} - 0 +  \frac{1}{\pi_c^3}E[(\hat{f} - \pi_c)^2] - \frac{1}{\pi_c^4}E[(\hat{f} - \pi_c)^3 ]+ \frac{1}{\pi_c^5}E[(\hat{f} - \pi_c)^4]\\
& = \frac{1}{\pi_c}+  \frac{\text{var}(\hat{f})}{\pi_c^2}\left[1 - \frac{1}{\pi_c^2}\frac{(1-2\pi_c)(1-2p)}{N-2} + \frac{1}{\pi_c^3}c_1\right]
\end{align*}
where $c_1$ is a constant related to the kurtosis.

More details on the moments:
\begin{align*}
E[(\hat{f}-\pi_c)^2] &= \text{var}(\hat{f}) = \frac{p(1-p)\pi_c(1-\pi_c)}{N-1}
\end{align*}

\begin{align*}
E[(\hat{f}-\pi_c)^3] &= \left(\text{var}(\hat{f})\right)^{3/2}\frac{(N-2n_c)(\sqrt{N-1})(N-2pN)}{(N-2)\sqrt{pn n_c(N-n_c)(1-p)N}}\\
&= \text{var}(\hat{f})\frac{\sqrt{p(1-p)\pi_c(1-\pi_c)}}{\sqrt{N-1}}\frac{(1-2\pi_c)(\sqrt{N-1})(1-2p)}{(N-2)\sqrt{p(1-p) \pi_c(1-\pi_c)}}\\
&= \text{var}(\hat{f})\frac{(1-2\pi_c)(1-2p)}{(N-2)}
\end{align*}

\begin{align*}
E[(\hat{f}-\pi_c)^4]
 &=\text{var}(\hat{f})c_1
\end{align*}
where $c_1$ is some constant related to kurtosis of order $1/N$.

Then we have
\begin{align*}
&E\left[\widehat{\CACE}_{\text{IV}}\right] - \CACE\\
 &=  \frac{1}{1-p}\left(1  - E\left[\frac{1}{ \hat{f}}\right]E[\hat{f}]\right)\left( \overline{Y}_c(0)-\overline{Y}_n\right)\\
&\approx - \frac{\left( \overline{Y}_c(0)-\overline{Y}_n\right)}{1-p} \frac{\text{var}(\hat{f})}{\pi_c}\left[1 - \frac{1}{\pi_c^2}\frac{(1-2\pi_c)(1-2p)}{N-2}+ \frac{1}{\pi_c^3}c_1\right]\\
\end{align*}

\subsection{Two-sided noncompliance}\label{append:two_sided_bias}

Using a Taylor expansion:
\begin{align*}
E\left[\frac{\widehat{\ITT}}{\hat{f}}\right] & \approx \frac{\ITT}{\pi_c} + \frac{\text{var}(\hat{f})\ITT}{\pi_c^3} - \frac{\text{Cov}(\widehat{\ITT}, \hat{f})}{\pi_c^2}
\end{align*}

For $\text{var}(\hat{f})$ we have

\begin{align*}
\text{var}(\hat{f}) &= \frac{1}{pN}\frac{1}{N-1}\sum_{i=1}^N(D_i(1) - \overline{D}(1))^2 + \frac{1}{(1-p)N}\frac{1}{N-1}\sum_{i=1}^N(D_i(0) - \overline{D}(0))^2\\
&  \quad - \frac{1}{N}\frac{1}{N-1}\sum_{i=1}^N(D_i(1) - D_i(0) - \overline{D}(1) +\overline{D}(0))^2 \\
&= \frac{1}{N-1}\left(\frac{1}{p} \pi_n(\pi_c+\pi_a)+ \frac{1}{1-p}\pi_a(\pi_n + \pi_c) - \pi_c(\pi_a +\pi_n)\right) \\
&= \frac{1}{N-1}\left(\frac{\pi_n(1-\pi_n)}{p} + \frac{\pi_a(1-\pi_a)}{1-p} - \pi_c(1-\pi_c)\right),
\end{align*}

Now for $\text{Cov}(\widehat{\ITT}, \hat{f})$:
\begin{align*}
\text{Cov}(\widehat{\ITT}, \hat{f}) &= \underbrace{\frac{1}{Np}\frac{1}{N-1}\sum_{i=1}^N(Y_i(1) - \overline{Y}(1))(D_i(1)-\overline{D}(1))}_{\text{A}}\\
&\quad  + \underbrace{\frac{1}{N(1-p)}\frac{1}{N-1}\sum_{i=1}^N(Y_i(0)-\overline{Y}(0))(D_i(0)-\overline{D}(0))}_{\text{B}}\\
&\quad  - \underbrace{ \frac{1}{N}\frac{1}{N-1}\sum_{i=1}^N(Y_i(1) - Y_i(0)-\ITT)(D_i(1)  - D_i(0) - \pi_c)}_{\text{C}}
\end{align*}

First we have for A,
\begin{align*}
\text{A} &= \frac{1}{Np}\frac{1}{N-1}\Big[\pi_n\sum_{i: D_i(1) = 1, D_i(0)=0}(Y_i(1) - \overline{Y}(1)) + \pi_n\sum_{i: D_i(1) = 1, D_i(0)=1}(Y_i(1) - \overline{Y}(1))\\
&\qquad  \qquad \qquad   - (\pi_c + \pi_a)\sum_{i: D_i(1) = 0, D_i(0)=0}(Y_i(1) - \overline{Y}(1))\Big]\\
&= \frac{1}{Np}\frac{1}{N-1}\Big[\pi_nn_c(\overline{Y}_c(1) - \overline{Y}(1)) + \pi_nn_a(\overline{Y}_a(1) - \overline{Y}(1))  - (\pi_c + \pi_a)n_n(\overline{Y}_n(0) - \overline{Y}(1))\Big]\\
&= \frac{\pi_n}{p(N-1)}\Big[\pi_c(\overline{Y}_c(1) - \overline{Y}(1)) + \pi_a(\overline{Y}_a(1) - \overline{Y}(1))  - (\pi_c + \pi_a)(\overline{Y}_n(0) - \overline{Y}(1))\Big]\\
&= \frac{\pi_n}{p(N-1)}\Big[\pi_c(\overline{Y}_c(1) - \overline{Y}_n(0)) + \pi_a(\overline{Y}_a(1) - \overline{Y}_n(0))\Big].
\end{align*}

We can get a similar simplification for B:
\begin{align*}
\text{B} &= \frac{\pi_a}{(1-p)(N-1)}\left[\pi_c(\overline{Y}_a(1) - \overline{Y}_c(0)) + \pi_n(\overline{Y}_a(1) - \overline{Y}_n(0))\right].
\end{align*}

Now for C:
\begin{align*}
\text{C} &= \frac{1}{N}\frac{1}{N-1}\Big[(\pi_a+\pi_n)\sum_{i: D_i(1) = 1, D_i(0)=0}(\CACE_i - \ITT) - \pi_c\sum_{i: D_i(1) = 1, D_i(0)=1}(0-\ITT)\\
&\qquad  \qquad \qquad - \pi_c\sum_{i: D_i(1) = 0, D_i(0)=0}(0-\ITT)\Big]\\
 &= \frac{1}{N}\frac{1}{N-1}\Big[(\pi_a+\pi_n)n_c(\CACE - \ITT) + \pi_cn_a\ITT + \pi_cn_n\ITT\Big]\\
  &= \frac{\pi_c}{N-1}\Big[(\pi_a+\pi_n)(\CACE - \ITT) + \pi_a\ITT + \pi_n\ITT\Big]\\
    &= \frac{\pi_c(\pi_a + \pi_n)}{N-1}\CACE \\
    &= \frac{\pi_c(1-\pi_c)}{N-1}\CACE
\end{align*}

Putting it together, we have
\begin{align*}
\text{Cov}(\widehat{\ITT}, \hat{f}) &= \frac{\pi_n}{p(N-1)}\Big[\pi_c(\overline{Y}_c(1) - \overline{Y}_n(0)) + \pi_a(\overline{Y}_a(1) - \overline{Y}_n(0))\Big]\\
&\quad + \frac{\pi_a}{(1-p)(N-1)}\left[\pi_c(\overline{Y}_a(1) - \overline{Y}_c(0)) + \pi_n(\overline{Y}_a(1) - \overline{Y}_n(0))\right]\\
&\quad  - \frac{\pi_c(1-\pi_c)}{N-1}\CACE\\
&= \frac{\pi_n\pi_c}{p(N-1)}(\overline{Y}_c(1) - \overline{Y}_n(0)) + \frac{\pi_n\pi_a}{p(1-p)(N-1)}(\overline{Y}_a(1) - \overline{Y}_n(0))\\
&\quad + \frac{\pi_a\pi_c}{(1-p)(N-1)}(\overline{Y}_a(1) - \overline{Y}_c(0)) - \frac{\pi_c(1-\pi_c)}{N-1}\CACE
\end{align*}

We plug our expressions into our original expansion and rearrange as so:
\begin{align*}
E\left[\frac{\widehat{\ITT}}{\hat{f}}\right] - \CACE & \approx  \frac{\text{var}(\hat{f})\ITT}{\pi_c^3} - \frac{\text{Cov}(\widehat{\ITT}, \hat{f})}{\pi_c^2}\\
& = \frac{\CACE}{\pi_c^2}\left[\frac{1}{N-1}\left(\frac{\pi_n(1-\pi_n)}{p} + \frac{\pi_a(1-\pi_a)}{1-p} - \pi_c(1-\pi_c)\right)\right] \\
&\quad - \frac{1}{\pi_c^2}\Big[\frac{\pi_n\pi_c}{p(N-1)}(\overline{Y}_c(1) - \overline{Y}_n(0)) + \frac{\pi_n\pi_a}{p(1-p)(N-1)}(\overline{Y}_a(1) - \overline{Y}_n(0))\\
&\qquad \qquad + \frac{\pi_a\pi_c}{(1-p)(N-1)}(\overline{Y}_a(1) - \overline{Y}_c(0)) - \frac{\pi_c(1-\pi_c)}{N-1}\CACE\Big]\\
& = \frac{\CACE}{\pi_c^2}\left[\frac{1}{N-1}\left(\frac{\pi_n(1-\pi_n)}{p} + \frac{\pi_a(1-\pi_a)}{1-p} \right)\right] \\
&\quad - \frac{1}{\pi_c^2}\Big[\frac{\pi_n\pi_c}{p(N-1)}(\overline{Y}_c(1) - \overline{Y}_n(0)) + \frac{\pi_n\pi_a}{p(1-p)(N-1)}(\overline{Y}_a(1) - \overline{Y}_n(0))\\
&\qquad \qquad + \frac{\pi_a\pi_c}{(1-p)(N-1)}(\overline{Y}_a(1) - \overline{Y}_c(0))\Big]\\
& = \frac{1}{\pi_c^2(N-1)}\Bigg[\left(\frac{\pi_n(\pi_a + \pi_c)}{p} + \frac{\pi_a(\pi_n + \pi_c)}{1-p} \right)(\overline{Y}_c(1) - \overline{Y}_c(0)) \\
&\quad - \Big[\frac{\pi_n\pi_c}{p}(\overline{Y}_c(1) - \overline{Y}_n(0)) + \frac{\pi_n\pi_a}{p(1-p)}(\overline{Y}_a(1) - \overline{Y}_n(0)) + \frac{\pi_a\pi_c}{1-p}(\overline{Y}_a(1) - \overline{Y}_c(0))\Big]\Bigg]\\
& = \frac{1}{\pi_c^2(N-1)}\Bigg[ \frac{\pi_n\pi_c}{p}(\overline{Y}_c(1) - \overline{Y}_c(0) - \overline{Y}_c(1) + \overline{Y}_n(0)) \\
& \qquad \qquad \qquad \qquad+ \frac{\pi_a\pi_c}{1-p}(\overline{Y}_c(1) - \overline{Y}_c(0) -\overline{Y}_a(1) +\overline{Y}_c(0))\\
&\qquad \qquad \qquad \qquad + \left(\frac{\pi_n\pi_a}{p} + \frac{\pi_a\pi_n}{1-p} \right)(\overline{Y}_c(1) - \overline{Y}_c(0))- \frac{\pi_n\pi_a}{p(1-p)}(\overline{Y}_a(1) - \overline{Y}_n(0)) \Bigg]\\
& = \frac{1}{\pi_c^2(N-1)}\Bigg[ \frac{\pi_n\pi_c}{p}(\overline{Y}_n(0) - \overline{Y}_c(0)) + \frac{\pi_a\pi_c}{1-p}(\overline{Y}_c(1) -\overline{Y}_a(1) )\\
& \qquad \qquad+ \frac{\pi_n\pi_a}{p(1-p)}(\overline{Y}_c(1) - \overline{Y}_c(0)  - \overline{Y}_a(1) +  \overline{Y}_n(0)) \Bigg]\\
& = \frac{1}{\pi_c^2(N-1)}\Bigg[ \frac{\pi_n((1-p)\pi_c + \pi_a)}{p(1-p)}(\overline{Y}_n(0) - \overline{Y}_c(0)) + \frac{\pi_a(p\pi_c + \pi_n)}{p(1-p)}(\overline{Y}_c(1) -\overline{Y}_a(1) )\Bigg]
\end{align*}

\section{Further details on and results for simulations}
\label{app:simulation}

\subsection{Simulation Design}
We generate data by first generating a four-category categorical covariate, $X_i$, to divide the units into strata.
Each stratum is then given a baseline control-side mean.
If our covariate is predictive of outcome, these means will vary; otherwise they are shared.

We next generate compliance behavior for each unit, flipping an independent coin with probability $p_i$, where $p_i$ depends on $X_i$ in the case of a covariate predictive of compliance, and $p_i$ is constant across units if not.
We keep overall compliance rate the same in either case.

After we have our $X_i$ and $S_i(0), S_i(1)$ (a pair of indicator variables indicating treatment take-up depending on treatment assignment), we generate our potential outcomes.
We assign a constant shift in the mean for never-takers.
For simplicity, we do not shift the always-takers, but our code is available for use and the option is there.
The potential outcomes are then generated as normal around these means, with a standard deviation set to achieve an overall (cross-strata) variance of 1.
This implies that if we do not have a prognostic $X_i$, the within-strata variance will be higher as total variance is within plus between variation.
The treatment effect is then added to the compliers' $Y_i(0)$ to get the $Y_i(1)$.

Once the dataset is generated, we randomize to treatment and control, and calculate $Y_i^{obs}$, the observed outcome.
We can then estimate the overall CACE with our different methods.

To illustrate trends, we selected parameters in our simulation such that when $X_i$ is predictive of something, it is very predictive.
For example, when $X_i$ predicts compliance, the top tier often has above 50\% compliance and the bottom tier usually has below 1\% compliance.
The average $R^2$ of the outcome regressed onto $X_i$ for the control group, for predictive $X_i$, is about 63\%.

For the auxiliary simulation to study covariates predictive of compliance, we set a tuning parameter $r$ from $[0, 1]$, and set the compliance of the four strata to $p, pr, pr^2$, and $pr^3$.
We then set $p$, a scaling factor, such that overall compliance equaled our target compliance rate $P$, using:
$$ P = w_1 pr^3 + w_2 pr^2 + w_3 pr + w_4 p , $$
where $w_k$ is the proportion of units in stratum $k$.
When $r=0$, all our compliers are in the last stratum.
When $r=1$, all strata have the same compliance rate.

\subsection{Further results on SE estimator performance}

\begin{figure}[hbt]
\center
  \includegraphics{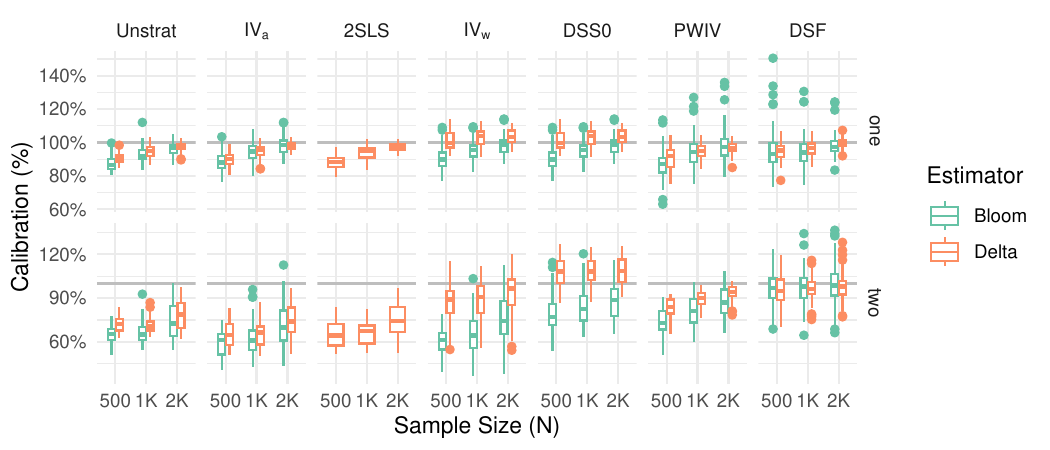}
  \caption{Relative percent change of \emph{median} estimated standard error to true standard error, for both the delta method and Bloom standard errors. Each point is a specific simulation scenario. Points above 100\% indicate standard errors systematically too large, and below systematically too small.}
  \label{fig:se_estimator_plot_median}
\end{figure}

In the main paper, we look at the average SE estimate, after they had been windsorized. The overall averages are still driven by the extreme outliers (10 standard deviations is still very large compared to typical values).
As a point of comparison, we have, analogous to Figure 4 in the main paper, the ratio of the \emph{median} estimated standard error to the true standard error on Figure~\ref{fig:se_estimator_plot_median}.
Here the true standard error is over the windorized estimates, but the point estimates are less frequently extreme than the standard error estimates.
Overall, we see that given the right skew in the SE estimates, the median estimated standard error tends to be too small, relative to the truth (even considering that the true SEs are too small given Windorization).

\subsection{Stability of SE Estimators}
\label{sec:se_stability}

Here we focus on one-sided noncompliance only, as two-sided noncompliance was clearly very unstable with large numbers of extreme point estimates and standard error estimates.
For one-sided noncompliance, we wanted to investigate whether the uncertainty of the stratified standard error estimators is generally lower, relative to the corresponding true standard error, as compared to the unstratified estimator.
For each estimator and simulation context we calculate the standard deviation of the estimated standard errors and divide by the Monte-Carlo estimated true standard error to obtain a relative average percent error in the uncertainty measure.
We then compare these ratios, for the post-stratified estimators, to the corresponding ratio of the baseline unstratified estimator.
This is a ratio of ratios:
\begin{equation*}
\mbox{relative SE}_{est} = \frac{sd( \widehat{SE}(\hat{\tau}_{est} )) }{ SE(\hat{\tau}_{est} ) } \slash \frac{ sd( \widehat{SE}(\hat{\tau}_{u} ) )}{ SE(\hat{\tau}_{u} ) },
\end{equation*}
where $est$ is an estimator of interest and $u$ is our baseline unstratified estimator.

The Bloom estimated standard errors for either stratified estimator are neither more or less unstable to any substantial degree, relative to their true precision, as compared to the unstratified.
For the delta method standard error, we do see higher relative instability in the standard error estimates for the within approach.
A subsequent analysis (not shown) shows that for a predictive covariate, even though the standard errors can be \emph{relatively} more uncertain, the overall reduction in uncertainty more than offsets this gain, resulting in a more precise estimate of uncertainty for a more precise estimator.
Without a predictive covariate, however, true precision gains are minimal, and the additional instability in estimating the standard errors does result in an overall cost.

The 2SLS standard error estimators are compared first to Bloom, then to delta standard error estimators for the unstratified IV estimate.
The splitting pattern shows that the relative stability of 2SLS standard error estimates are not as unstable as the delta method, but are less stable than Bloom.

\begin{figure}[hbt]
\center
  \includegraphics{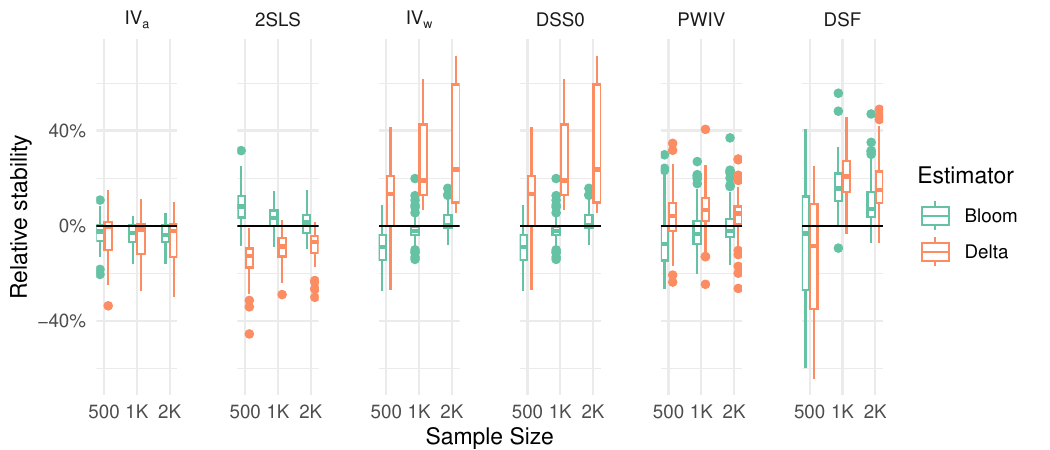}
  \caption{Ratio of relative instabilities of $\widehat{SE}$ (centered at 100\%), with instability taken as the ratio of the standard deviation of $\widehat{SE}$ vs. the true $SE$ (see text for equation), and the primary ratio being the post-stratified instability vs. unstratified instability. Numbers below 0 indicate the post-stratified standard error estimates are relatively more stable than unstratified standard error estimates, as indexed by their relative uncertainties.}
  \label{fig:se_instability_plot}
\end{figure}

\subsection{Random post-stratification}
\label{app:random_poststrat}

\begin{figure}[hbt]
\center
  \includegraphics{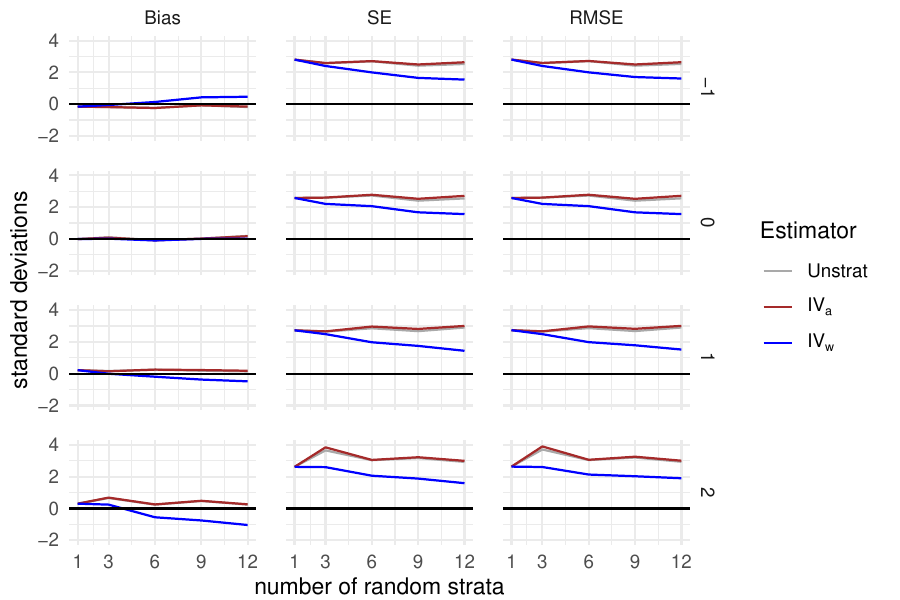}
  \caption{Bias, SE, and RMSE when randomly stratifying units with varying number of strata ($x$-axis) and varying amount of separation between the never-takers and compliers (rows of results) for one-sided noncompliance. SE gains swamp bias cost in these scenarios, even when there is a great deal of separation (2 standard deviations) between the means of the never-takers and compliers.}
  \label{fig:random_strat_nohet}
\end{figure}

We were surprised to see benefits to post-stratification when stratifying on a variable that is neither predictive of compliance status or outcome for one-sided noncompliance.
To further verify this finding, we conducted an additional simulation study where we first generated a dataset as we did for our primary simulation, and then generated a categorical covariate entirely at random to go with it.
We explored generating such a covariate with 1, 3, 6, 9, and 12 categories.
Results are on Figure~\ref{fig:random_strat_nohet}.
We also varied the extent to which the never-takers are systematically different from compliers.

We see benefits to stratification for $IV_{w}$, although bias does climb the more the never-takers are different from compliers.
What is happening is if we end up with a strata that has no compliers in treatment, that entire strata is dropped.
This reduces overall noise as we know that group does not provide any hope of a treatment by control comparison, as there is no information about compliers on the treatment side.
We do end up with bias since this will systematically drop strata with no treatment compliers, but not drop strata with no control compliers, creating systematic imbalance.
Even so, we see the gains from reduced instability offsets this bias in these scenarios.
This is perhaps more a statement about the instability of the overall IV estimator (note the SEs, in effect size units, are larger than 1 effect size unit), than small bias.

\subsection{Violations of the exclusion restriction}\label{supsubsec:er}

\begin{figure}[hbt]
\center
  \includegraphics{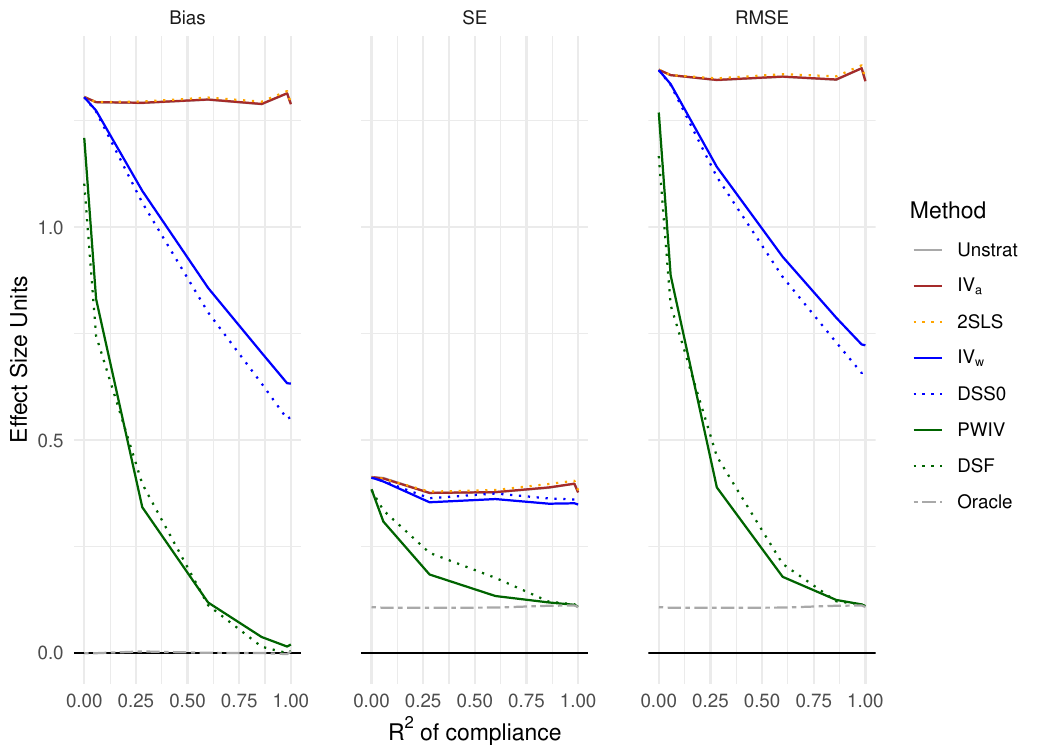}
  \caption{Bias, SE, and RMSE when exclusion restriction is violated.  As compliers get increasingly concentrated in fewer strata, the estimators that drop or down-weight low-complier weight strata have substantial bias reductions}
  \label{fig:exclusion_restriction}
\end{figure}

We modified the simulation scenario where we varied the concentration of the compliers in the last strata by simply adding an overall 0.20 effect size impact of treatment to all noncompliers, and a 0.50 effect size impact for the compliers.
Results are on Figure~\ref{fig:exclusion_restriction}.

First, we see a large bias, larger than the 0.20 impact.
This is because 85\% of the units are noncompliers, and all the ITT due to them gets attributed to the 15\% compliers.
Even mild violations of the exclusion restriction can be dangerous.

We also see that as we have a covariate increasingly predictive of complier status, we can carve out a subgroup that has a higher proportion of compliers, which mitigates the bias.

%% file: IV_poststratification-applied.bbl
\begin{thebibliography}{}

\bibitem[Angrist et~al., 1996]{AngImbRub96}
Angrist, J.~D., Imbens, G.~W., and Rubin, D.~B. (1996).
\newblock Identification of causal effects using instrumental variables.
\newblock {\em Journal of the American Statistical Association},
  91(434):444--455.

\bibitem[Arceneaux et~al., 2012]{arceneaux2012get}
Arceneaux, K., Kousser, T., and Mullin, M. (2012).
\newblock Get out the vote-by-mail? a randomized field experiment testing the
  effect of mobilization in traditional and vote-by-mail precincts.
\newblock {\em Political Research Quarterly}, 65(4):882--894.

\bibitem[Arceneaux and Nickerson, 2009]{Nickerson:2009}
Arceneaux, K. and Nickerson, D.~W. (2009).
\newblock Who is mobilized to vote? a re-analysis of 11 field experiments.
\newblock {\em American Journal of Political Science}, 52(1):1--16.

\bibitem[Baiocchi et~al., 2012]{baiocchi2012near}
Baiocchi, M., Small, D.~S., Yang, L., Polsky, D., and Groeneveld, P.~W. (2012).
\newblock Near/far matching: a study design approach to instrumental variables.
\newblock {\em Health Services and Outcomes Research Methodology},
  12(4):237--253.

\bibitem[Beach and Meier, 1989]{Beach:1989wr}
Beach, M.~L. and Meier, P. (1989).
\newblock {Choosing covariates in the analysis of clinical trials}.
\newblock {\em Controlled clinical trials}, 10(4):161 -- 175.

\bibitem[Blackwell and Pashley, 2023]{blackwellpashley}
Blackwell, M. and Pashley, N.~E. (2023).
\newblock Noncompliance and instrumental variables for 2k factorial
  experiments.
\newblock {\em Journal of the American Statistical Association},
  118(542):1102--1114.

\bibitem[Bloom, 1984]{Bloom:1984}
Bloom, H.~S. (1984).
\newblock Accounting for no-shows in experimental evaluation designs.
\newblock {\em Evaluation review}, 8(2):225--246.

\bibitem[Bound et~al., 1995]{Bound:1995}
Bound, J., Jaeger, D., and Baker, R. (1995).
\newblock Problems with intrustmental variables estimation when the correlation
  between the instruments and the endogenous explanatory variable is weak.
\newblock {\em Journal of the American Statistical Association},
  90(430):443--450.

\bibitem[Bowers et~al., 2008]{gg_gotv_data}
Bowers, J., Hansen, B.~B., and Fredrickson, M. (2008).
\newblock {Replication data for: Attributing Effects to A Cluster Randomized
  Get-Out-The-Vote Campaign: The Compendium}.

\bibitem[Coppock et~al., 2022]{coppock2022does}
Coppock, A., Green, D.~P., and Porter, E. (2022).
\newblock Does digital advertising affect vote choice? evidence from a
  randomized field experiment.
\newblock {\em Research \& Politics}, 9(1):20531680221076901.

\bibitem[Ding and Lu, 2017]{ding2017principal}
Ding, P. and Lu, J. (2017).
\newblock Principal stratification analysis using principal scores.
\newblock {\em Journal of the Royal Statistical Society: Series B (Statistical
  Methodology)}, 79(3):757--777.

\bibitem[Feller et~al., 2017]{feller2017principal}
Feller, A., Mealli, F., and Miratrix, L. (2017).
\newblock Principal score methods: Assumptions, extensions, and practical
  considerations.
\newblock {\em Journal of Educational and Behavioral Statistics},
  42(6):726--758.

\bibitem[Frangakis and Rubin, 2002]{Frangakis:2002}
Frangakis, C.~E. and Rubin, D.~B. (2002).
\newblock {Principal Stratification in Causal Inference}.
\newblock {\em Biometrics,}, 58(1):21 -- 29.

\bibitem[Gerber and Green, 2000]{Gerber:2000}
Gerber, A.~S. and Green, D.~P. (2000).
\newblock The effects of personal canvassing, telephone calls, and direct mail
  on voter turnout: A field experiment.
\newblock {\em American Political Science Review}, 94(3):653--663.

\bibitem[Gerber and Green, 2012]{gerber2012field}
Gerber, A.~S. and Green, D.~P. (2012).
\newblock {\em Field experiments: Design, analysis, and interpretation}.
\newblock WW Norton.

\bibitem[Gerber et~al., 2008]{Gerber:2008}
Gerber, A.~S., Green, D.~P., and Larimer, C.~W. (2008).
\newblock Social pressure and voter turnout: Evidence from a large-scale field
  experiment.
\newblock {\em American Political Science Review}, 102(1):33--48.

\bibitem[Green and Gerber, 2016]{green2016voter}
Green, D.~P. and Gerber, A.~S. (2016).
\newblock Voter mobilization, experimentation, and translational social
  science.
\newblock {\em Perspectives on Politics}, 14(3):738--749.

\bibitem[Green and Gerber, 2019]{green2019get}
Green, D.~P. and Gerber, A.~S. (2019).
\newblock {\em Get out the vote: How to increase voter turnout}.
\newblock Brookings Institution Press.

\bibitem[Green et~al., 2003a]{Green:2003a}
Green, D.~P., Gerber, A.~S., and Nickerson, D.~W. (2003a).
\newblock Getting out the vote in local elections: Results from six
  door-to-door canvassing experiments.
\newblock {\em Journal of Politics}, 65(4):1083--1096.

\bibitem[Green et~al., 2003b]{gotv_data1}
Green, D.~P., Gerber, A.~S., and Nickerson, D.~W. (2003b).
\newblock {Replication Materials for: Getting Out the Vote in Local Elections:
  Results from Six Door-to-Door Canvassing Experiments}.

\bibitem[Green et~al., 2013]{green2013field}
Green, D.~P., McGrath, M.~C., and Aronow, P.~M. (2013).
\newblock Field experiments and the study of voter turnout.
\newblock {\em Journal of Elections, Public Opinion and Parties}, 23(1):27--48.

\bibitem[Green and Zelizer, 2017]{green2017much}
Green, D.~P. and Zelizer, A. (2017).
\newblock How much gotv mail is too much? results from a large-scale field
  experiment.
\newblock {\em Journal of Experimental Political Science}, 4(2):107--118.

\bibitem[Guryan et~al., 2023]{guryan2023not}
Guryan, J., Ludwig, J., Bhatt, M.~P., Cook, P.~J., Davis, J.~M., Dodge, K.,
  Farkas, G., Fryer~Jr, R.~G., Mayer, S., Pollack, H., et~al. (2023).
\newblock Not too late: Improving academic outcomes among adolescents.
\newblock {\em American Economic Review}, 113(3):738--765.

\bibitem[Hansen and Bowers, 2009]{hansen2009attributing}
Hansen, B.~B. and Bowers, J. (2009).
\newblock Attributing effects to a cluster-randomized get-out-the-vote
  campaign.
\newblock {\em Journal of the American Statistical Association},
  104(487):873--885.

\bibitem[Heller, 2014]{heller2014summer}
Heller, S.~B. (2014).
\newblock Summer jobs reduce violence among disadvantaged youth.
\newblock {\em Science}, 346(6214):1219--1223.

\bibitem[Hern\'{a}n and Robins, 2006]{Hernan:2006}
Hern\'{a}n, M.~A. and Robins, J.~M. (2006).
\newblock Instruments for causal inference: An epidemiologists dream.
\newblock {\em Epidemiology}, 17(4):360--372.

\bibitem[Hernan and Robins, 2019]{hernan2019}
Hernan, M.~A. and Robins, J.~M. (2019).
\newblock {\em Causal inference}.
\newblock CRC Boca Raton, forthcoming.

\bibitem[Hirano et~al., 2000]{hirano2000assessing}
Hirano, K., Imbens, G.~W., Rubin, D.~B., and Zhou, X.-H. (2000).
\newblock Assessing the effect of an influenza vaccine in an encouragement
  design.
\newblock {\em Biostatistics}, 1(1):69--88.

\bibitem[Imbens and Rubin, 1997]{imbens1997bayesian}
Imbens, G.~W. and Rubin, D.~B. (1997).
\newblock Bayesian inference for causal effects in randomized experiments with
  noncompliance.
\newblock {\em The annals of statistics}, pages 305--327.

\bibitem[Imbens and Rubin, 2015]{CausalInferenceText}
Imbens, G.~W. and Rubin, D.~B. (2015).
\newblock {\em Causal Inference for Statistics, Social, and Biomedical
  Sciences: An Introduction}.
\newblock Cambridge University Press, New York.

\bibitem[Kang et~al., 2018]{Keele:2017fiv}
Kang, H., Peck, L., and Keele, L. (2018).
\newblock Inference for instrumental variables: A randomization inference
  approach.
\newblock {\em Journal of The Royal Statistical Society, Series A},
  181(4):1131--1154.

\bibitem[Keele and Morgan, 2016]{keele2016strong}
Keele, L. and Morgan, J.~W. (2016).
\newblock How strong is strong enough? strengthening instruments through
  matching and weak instrument tests.
\newblock {\em The Annals of Applied Statistics}, 10(2):1086--1106.

\bibitem[Kleiber and Zeileis, 2008]{aer}
Kleiber, C. and Zeileis, A. (2008).
\newblock {\em Applied Econometrics with {R}}.
\newblock Springer-Verlag, New York.
\newblock {ISBN} 978-0-387-77316-2.

\bibitem[Li and Ding, 2017]{LiDin17}
Li, X. and Ding, P. (2017).
\newblock General forms of finite population central limit theorems with
  applications to causal inference.
\newblock {\em Journal of the American Statistical Association},
  112(520):1759--1769.

\bibitem[Mann et~al., 2020]{mann2020negatively}
Mann, C.~B., Arceneaux, K., and Nickerson, D.~W. (2020).
\newblock Do negatively framed messages motivate political participation?
  evidence from four field experiments.
\newblock {\em American Politics Research}, 48(1):3--21.

\bibitem[Mealli and Mattei, 2012]{mealli2012refreshing}
Mealli, F. and Mattei, A. (2012).
\newblock A refreshing account of principal stratification.
\newblock {\em The international journal of biostatistics}, 8(1).

\bibitem[Mealli and Pacini, 2013]{mealli2013using}
Mealli, F. and Pacini, B. (2013).
\newblock Using secondary outcomes to sharpen inference in randomized
  experiments with noncompliance.
\newblock {\em Journal of the American Statistical Association},
  108(503):1120--1131.

\bibitem[Miratrix et~al., 2018]{miratrix2018bounding}
Miratrix, L., Furey, J., Feller, A., Grindal, T., and Page, L.~C. (2018).
\newblock Bounding, an accessible method for estimating principal causal
  effects, examined and explained.
\newblock {\em Journal of Research on Educational Effectiveness},
  11(1):133--162.

\bibitem[Miratrix et~al., 2013]{miratrix2013adjusting}
Miratrix, L.~W., Sekhon, J.~S., and Yu, B. (2013).
\newblock Adjusting treatment effect estimates by post-stratification in
  randomized experiments.
\newblock {\em Journal of the Royal Statistical Society: Series B (Statistical
  Methodology)}, 75(2):369--396.

\bibitem[Nickerson, 2006]{Nickerson:2006}
Nickerson, D.~W. (2006).
\newblock Volunteer phone calls can increase turnout.
\newblock {\em American Politics Research}, 34(3):271--292.

\bibitem[Pashley, 2022]{pashley2019note}
Pashley, N.~E. (2022).
\newblock Note on the delta method for finite population inference with
  applications to causal inference.
\newblock {\em Statistics \& Probability Letters}, 188:109540.

\bibitem[Pashley and Miratrix, 2021]{pashley2021insights}
Pashley, N.~E. and Miratrix, L.~W. (2021).
\newblock Insights on variance estimation for blocked and matched pairs
  designs.
\newblock {\em Journal of Educational and Behavioral Statistics},
  46(3):271--296.

\bibitem[Pashley and Miratrix, 2022]{pashley2020block}
Pashley, N.~E. and Miratrix, L.~W. (2022).
\newblock Block what you can, except when you shouldn't.
\newblock {\em Journal of Educational and Behavioral Statistics},
  47(1):69--100.

\bibitem[Rubin, 1980]{rubin_1980}
Rubin, D.~B. (1980).
\newblock Randomization analysis of experimental data: {T}he {F}isher
  randomization test comment.
\newblock {\em J. Amer. Statist. Assoc.}, 75(371):591--593.

\bibitem[Schochet, 2023]{schochet2023design}
Schochet, P.~Z. (2023).
\newblock Design-based rct estimators and central limit theorems for baseline
  subgroup and related analyses.
\newblock {\em arXiv preprint arXiv:2310.08726}.

\bibitem[Schochet, 2024]{schochet2024design}
Schochet, P.~Z. (2024).
\newblock Design-based estimation and central limit theorems for local average
  treatment effects for rcts.
\newblock {\em arXiv preprint arXiv:2401.07401}.

\bibitem[Senn, 1989]{Senn:1989wk}
Senn, S.~J. (1989).
\newblock {Covariate imbalance and random allocation in clinical trials}.
\newblock {\em Statistics in Medicine}, 8(4):467 -- 475.

\bibitem[Small and Rosenbaum, 2008]{Small:2008}
Small, D.~S. and Rosenbaum, P.~R. (2008).
\newblock War and wages: the strength of instrumental variables and their
  sensitivity to unobserved biases.
\newblock {\em Journal of the American Statistical Association},
  103(483):924--933.

\bibitem[Splawa-Neyman et~al., 1990]{SplawaNeyman:1990ux}
Splawa-Neyman, J., Dabrowska, D.~M., and Speed, T.~P. (1990).
\newblock {On the application of probability theory to agricultural
  experiments. Essay on principles. Section 9}.
\newblock {\em Statistical Science}, 5(4):465 -- 472.

\bibitem[Staiger and Stock, 1997]{Staiger:1997}
Staiger, D. and Stock, J.~H. (1997).
\newblock Instrumental variables regression with weak instruments.
\newblock {\em Econometrica}, 65:557--586.

\bibitem[Stock and Yogo, 2005]{Stock:2005}
Stock, J.~H. and Yogo, M. (2005).
\newblock Testing for weak instruments in linear iv regression.
\newblock In Andrews, D.~W. and Stock, J.~H., editors, {\em Identification and
  Inference in Econometric Models: Essays in Honor of Thomas J. Rothenberg},
  chapter~5. Cambridge University Press.

\bibitem[Ten~Have et~al., 2004]{ten2004causal}
Ten~Have, T.~R., Elliott, M.~R., Joffe, M., Zanutto, E., and Datto, C. (2004).
\newblock Causal models for randomized physician encouragement trials in
  treating primary care depression.
\newblock {\em Journal of the American Statistical Association},
  99(465):16--25.

\bibitem[Wang and Tchetgen~Tchetgen, 2018]{wang2018}
Wang, L. and Tchetgen~Tchetgen, E. (2018).
\newblock Bounded, efficient and multiply robust estimation of average
  treatment effects using instrumental variables.
\newblock {\em Journal of the Royal Statistical Society: Series B (Statistical
  Methodology)}, 80(3):531--550.

\end{thebibliography}
